\documentclass[twocolumn]{aastex63}

\graphicspath{{./}{figures1/}}

\received{2019 October 2}
\revised{2020 January 16}
\accepted{2020 January 17}

\submitjournal{ApJ}
\usepackage{amsmath}
\usepackage{chemformula}
\let\ce\ch
\usepackage{makecell}

\shorttitle{\texttt{Helios-r2}}
\shortauthors{Kitzmann et al.}

\begin{document}

\title{\texttt{Helios-r2} - A New Bayesian, Open-source Retrieval Model for Brown Dwarfs and Exoplanet Atmospheres}

\correspondingauthor{Daniel Kitzmann, Kevin Heng}
\email{daniel.kitzmann@csh.unibe.ch, kevin.heng@csh.unibe.ch}
\author[0000-0003-4269-3311]{Daniel Kitzmann}
\affil{Center for Space and Habitability, University of Bern, Gesellschaftsstrasse 6, Bern, 3012, Switzerland}

\author[0000-0003-1907-5910]{Kevin Heng}
\affil{Center for Space and Habitability, University of Bern, Gesellschaftsstrasse 6, Bern, 3012, Switzerland}

\author[0000-0002-3327-1072]{Maria Oreshenko}
\affil{Center for Space and Habitability, University of Bern, Gesellschaftsstrasse 6, Bern, 3012, Switzerland}

\author[0000-0002-0632-4407]{Simon L. Grimm}
\affil{Center for Space and Habitability, University of Bern, Gesellschaftsstrasse 6, Bern, 3012, Switzerland}

\author[0000-0003-3714-5855]{D\'{a}niel Apai}
\affil{Steward Observatory, The University of Arizona, Tucson, AZ 85721, USA}
\affil{Lunar and Planetary Laboratory, The University of Arizona, Tucson, AZ 85721, USA}

\author[0000-0003-2649-2288]{Brendan P. Bowler}
\affil{McDonald Observatory and the University of Texas at Austin \\
	Department of Astronomy, 2515 Speedway, Stop C1400, Austin, TX 78712, USA}

\author[0000-0002-6523-9536]{Adam J. Burgasser}
\affil{Center for Astrophysics and Space Science, University of California San Diego, La Jolla, CA 92093, USA}

\author[0000-0002-5251-2943]{Mark S. Marley}
\affil{ NASA Ames Research Center Moffett Field, Mountain View, CA 94035, USA}

\begin{abstract}

We present an improved, hybrid CPU-GPU atmospheric retrieval code, \texttt{Helios-r2}, which is applicable to medium-resolution emission spectra of brown dwarfs, in preparation for precision atmospheric spectroscopy in the era of the \textit{James Webb Space Telescope}. The model is available as open-source code on the Exoclimes Simulation Platform.
We subject \texttt{Helios-r2} to a battery of tests of varying difficulty.  The simplest test involves a mock retrieval on a forward model generated using the same radiative transfer technique, the same implementation of opacities, and the same chemistry model.  The least trivial test involves a mock retrieval on synthetic spectra from the \texttt{Sonora} model grid, which uses a different radiative transfer technique, a different implementation of opacities, and a different chemistry model.  A calibration factor, which is included to capture uncertainties in the brown dwarf radius, distance to the brown dwarf and flux calibration of the spectrum, may compensate, sometimes erroneously, for discrepancies in modeling choices and implementation.  We analyze spectra of the benchmark brown dwarf GJ 570 D and the binary brown dwarf companions in the Epsilon Indi system. The retrieved surface gravities are consistent with previous studies and/or values inferred from dynamical masses (for Epsilon Indi Ba and Bb only).  There remains no clear criterion on how to reject unphysical values of the retrieved brown dwarf radii. The inferred radii and corresponding masses should be taken with great caution. The retrieved carbon-to-oxygen ratios and metallicity depend on whether chemical equilibrium is assumed.  

\end{abstract}


\section{Introduction} \label{sec:intro}

Atmospheric retrieval solves the inverse problem of inferring the properties of an atmosphere given an emission or transmission spectrum of an exoplanet.  It has a rich legacy from the Earth remote sensing \citep[e.g.][]{Rodgers2000imas.book.....R} and planetary science \citep[e.g.][]{Irwin2008JQSRT.109.1136I} communities.  Early efforts focused on low-resolution spectra from transiting \citep[e.g.][]{Madhusudhan2009ApJ...707...24M, Line2012ApJ...749...93L} and directly imaged exoplanets \citep[e.g.][]{Lee2013ApJ...778...97L, Lavie2017AJ....154...91L}. When interpreted within a Bayesian framework \citep[e.g.][]{Benneke2012ApJ...753..100B}, the interpretation of low-resolution spectra is somewhat degenerate \citep[e.g.][]{Fisher2018MNRAS.481.4698F, Fisher2019ApJ...881...25F}.  With the upcoming \textit{James Webb Space Telescope} (\textit{JWST}), scheduled for launch in 2021, the exoplanet community is anticipating a transformational leap from low-to medium-resolution spectra (resolution $\sim$100--1000) across a broad wavelength range (0.6$-$28 $\mu$m).  While individual spectral lines will not be resolved, the shapes of families of lines will be accurately measured, which will break degeneracies  \citep[e.g.][]{Fisher2018MNRAS.481.4698F}. The broad wavelength coverage of the spectral continuum will enable constraints on the properties of aerosols, clouds, and hazes to be set \citep[e.g.][]{Kitzmann2018MNRAS.475...94K}.  

In the \textit{JWST} data regime, it is anticipated that details such as the parameterization of the temperature--pressure profile, clouds, and chemistry will become important. Furthermore, retrieval codes constructed by different research groups use different implementations of radiative transfer techniques and opacities.  While these details may not strongly affect the interpretation of low-resolution spectra, it is anticipated that they will lead to non-trivial differences in the interpretation of JWST spectra. This has, for example, been demonstrated by \citet{Rocchetto2016ApJ...833..120R} who studied the impact of the parameterization of the temperature--pressure profile retrievals of JWST-like spectra. The current study is the first in a series of papers that introduces a next-generation atmospheric retrieval code constructed with these details in mind.

Spectra of brown dwarfs provide an important testbed during this transition period between the \textit{Hubble Space Telescope} and \textit{JWST}. There are currently many more high-quality spectra available for brown dwarfs than exoplanets. For example, the SpeX Prism Library \citep{2014ASInC..11....7B} is a public repository\footnote{\url{http://pono.ucsd.edu/~adam/browndwarfs/spexprism/}} containing hundreds of low-resolution ($\lambda/\Delta\lambda$ $\approx$ 100), near-infrared (0.8--2.4~$\mu$m) brown dwarf spectra obtained with the SpeX near-infrared spectrograph \citep{2003PASP..115..362R} on the NASA Infrared Telescope Facility. Other large spectral datasets have been compiled from observations obtained with Keck/NIRSPEC \citep{2017ApJ...838...73M}, Hubble Space Telescope/WFC3 \citep[]{Manjavacas2019}, and multiple instruments \citep{2018csss.confE..51C}. The interpretation of brown dwarf spectra faces the same challenges as those of directly imaged exoplanets: generally, the radii and masses are unknown, which introduces degeneracies into the retrieval outcome.  Furthermore, the desire to retrieve the carbon-to-oxygen (C/O) ratio and metallicity hinges on whether all of the carbon- and oxygen-bearing molecules have been robustly detected and how to translate the retrieved elemental abundances of carbon and oxygen into those of more refractory elements, such as iron.

For the current study, we select three brown dwarfs as case studies. The first is the benchmark brown dwarf \object{GJ 570 D} \citep{Burgasser2000ApJ...531L..57B}, which has previously been studied using a retrieval model by \cite{Line2015ApJ...807..183L}.  The second and third are \object{Epsilon Indi Ba} and Bb \citep{King2010AA...510A..99K}, which are brown dwarfs in a binary system with measured dynamical masses \citep{dieterich18}.  GJ~570~D allows us to compare our retrieval outcomes to a string of previous studies, whereas $\epsilon$ Indi~Ba and Bb allow us to confront our retrieved gravities with those estimated from the dynamical masses.

In Section~\ref{sect:forward_model}, we describe the forward model of \texttt{Helios-r2}, including the radiative transfer technique used, our implementation of the atmospheric opacities, the chemistry model, and a novel approach to parameterizing the temperature--pressure profile using finite elements. Section~\ref{sect:nested_sampling} describes our implementation of the nested sampling method. \texttt{Helios-r2} is subjected to a battery of tests of varying difficulty in Section~\ref{sect:tests} and applied to the three case studies in Section~\ref{sec:brown_dwarf_retrieval}.  Discussion and summary are found in Sections \ref{sect:discussion} and \ref{sect:summary}, respectively.

The model is available as open-source software on the Exoclimes Simulation Platform\footnote{\label{fnote:EEG}\url{http://www.exoclime.org/}}. The EEG additionally offers a variety of other codes devoted to model and study atmospheres of planets and stars, including a general circulation model (\texttt{THOR}, \citet{Mendonca2016ApJ...829..115M}) and the ultra-fast equilibrium chemistry \texttt{FastChem} \citep{Stock2018MNRAS.479..865S}, as well as an alternative retrieval approach based on the random forest machine-learning technique (\texttt{HELA}, \citet{MarquezNeila018NatAs...2..719M}). The general radiation package is referred to as \texttt{Helios} and consists of the opacity calculator \texttt{Helios-k} \citep{Grimm2015ApJ...808..182G}, the radiative transfer model \texttt{Helios} \citep{Malik2017AJ....153...56M} and \texttt{Helios-r2}.

\section{Forward Model}
\label{sect:forward_model}
In this stud,y we use our newly developed retrieval model \texttt{Helios-r2} \citep{Kitzmann_Helios-r2_09}. This model is a complete rewrite of the original \texttt{Helios-r} retrieval model \citep{Lavie2017AJ....154...91L} and is available as open-source code on the Exoclimes Simulation Platform.

\texttt{Helios-r2} has been specifically adapted to describe atmospheres of brown dwarfs. It is a one-dimensional, semi-infinite atmosphere model that, for a given set of parameters, calculates the spectrum of a brown dwarf. Since the model has to be run within a Bayesian nested sampling approach, it has to be computationally as fast as possible. In contrast to a fully self-consistent model of such an atmosphere, we therefore have to apply a number of approximations and simplifications.

\texttt{Helios-r2} is programmed in standard C\texttt{++} and uses NVIDIA's CUDA language to execute the computationally heavy part of the forward model on a graphics card (GPU). It can run on both a pure CPU setup or a combination of CPUs and GPUs.

To test its applicability, we apply our new retrieval model on T spectral type brown dwarfs in this study. Later T dwarfs are usually well described by cloud-free models. In the present forward model of \texttt{Helios-r2}, clouds are therefore neglected. The one-dimensional atmosphere is partitioned into a number of levels/layers, distributed equidistantly in $\log p$-space. We use 70 levels (i.e., 69 layers) throughout the study. Increasing this number further has proven to have no effect on the retrieval results. In the following subsections, we provide additional details on, for example, the radiative transfer, the opacity sources, the chemistry, and the description of the temperature profile.

\subsection{Radiative Transfer}

For a given source function $S_\nu$, the radiative transfer equation in a plane-parallel, semi-infinite atmosphere has a simple solution \citep[e.g.][]{Mihalas1978stat.book.....M}, given by
\begin{equation}
  I_\nu^+(\tau_\nu, \mu) =  \int_{\tau_\nu}^{\infty} S_\nu(t)e^{-(t-\tau_\nu)/\mu} \mathrm d t /\mu \ ,
  \label{eq:rt_formal_solution}
\end{equation}
where $I_\nu^+$ is the outgoing intensity, i.e., for $0 < \mu \leq 1$. In the following, we neglect scattering. Thus, the source function is simply given by
\begin{equation}
S_\nu (\tau_\nu) = B_\nu(T(\tau_\nu)) ,
\end{equation}
where $B_\nu(T(\tau_\nu))$ is the Planck function with a temperature $T$ at a given optical depth $\tau_\nu$.
Equation \eqref{eq:rt_formal_solution} can be formally integrated with respect to the angular variable $\mu$ to yield the angular moments of the radiation field, such as the mean intensity or the flux. The resulting equations are known as the Schwarzschild--Milne equations and for the outgoing flux $F_\nu^+$ given by
\begin{equation}
  F_\nu^+(\tau_\nu) = 2 \pi \int_{\tau_\nu}^{\infty} S_\nu(t_\nu) E_2(t_\nu - \tau_\nu) \mathrm d t_\nu \ ,
  \label{eq:flux_schwarzschild-milne}
\end{equation}
where $E_2$ is the second exponential integral.

While it is possible to directly integrate Eq.~ \eqref{eq:flux_schwarzschild-milne} to obtain the outgoing flux at the upper atmosphere ($\tau_\nu = 0$), this approach might lead to numerical and computational difficulties. Evaluating the exponential integrals can be costly in terms of computational time and becomes unstable at high optical depths. Additionally, numerically integrating the equation by using the trapezoidal rule can lead to rather large numerical errors that accumulate along the path of integration unless a high vertical resolution is used.

To circumvent these problems, we instead employ the so-called short characteristics method here, first described by \citet{Olson1987JQSRT..38..325O}. Essentially, this method solves the characteristic of the radiative transfer equation on a layer-by-layer basis. Additionally, to stabilize the integration, a weighting function of the form $e^{-\tau/ \mu}$ is introduced.
For a single layer, Eq. \eqref{eq:rt_formal_solution} can be written as
\begin{equation}
  I_\nu^+(\tau_{\nu,i}, \mu) = I_\nu^+(\tau_{\nu,i-1}, \mu) e^{-\tau_{\nu,i}} + \Delta I_{\nu,i}^+(S_\nu, \mu) \ ,
  \label{eq:rt_short_char}
\end{equation}
with
\begin{equation}
  \Delta I_{\nu,i}^+(S_\nu, \mu) = \alpha_i S_{i+1} + \beta_i S_i + \gamma_i S_{i-1} \ .
\end{equation}
Here, $\tau_{\nu,i}$ refers to the optical depth in the $i$th layer, while $\alpha$, $\beta$, and $\gamma$ are coupling coefficients that connect the adjacent layers. Without scattering, the coefficient $\alpha$ is zero for outgoing rays. By assuming that the source function varies linearly within the layer, the coefficients are given by \citep{Olson1987JQSRT..38..325O}
\begin{equation}
  \beta_i = 1 + \frac{e^{-\Delta} - 1}{\Delta} \quad
  \gamma_i = e^{-\Delta} - \frac{e^{-\Delta} - 1}{\Delta} \ ,
\end{equation}
with
\begin{equation}
  \Delta = \frac{\tau_i - \tau_{i-1}}{\mu} \ .
\end{equation}
It is also possible to use higher-order interpolants of the source function. For the parabolic case, the coefficients can be found in \citet{Olson1987JQSRT..38..325O}. While offering higher accuracy per layer, these coefficients are also computationally more expensive.

We solve Eq.~\eqref{eq:rt_short_char} for a set of angles $\mu$, distributed according to a Gau{\ss} quadrature scheme \citep{Gauss1814} and then numerically integrate the intensities to obtain the outgoing flux
\begin{equation}
  F_\nu^+ = 2 \pi \int_0^1 I_\nu(\mu) \mu \, \mathrm d \mu \ .
  \label{eq:flux_model}
\end{equation}
For this study, we use two angles in the upward direction. This is equivalent to a four-stream discrete ordinate radiative transfer, which is usually sufficient for a problem without scattering and an atmosphere where the scale height is much smaller than the planet's radius. Compared to the direct solution of Eq.~ \eqref{eq:flux_schwarzschild-milne}, this method only requires the evaluation of two exponentials per layer.

The CPU part of \texttt{Helios-r2} is also equipped with the radiative transfer library CDISORT \citep{Stamnes1988ApOpt, Hamre2013AIPC.1531..923H}. This radiative transfer model uses a complex, multi-stream approach to solve the radiative transfer equation and provides the exact solution to the problem if the number of streams is large enough. While this library is usually too slow to run within a retrieval, we use it to verify the accuracy of our implementation of the short characteristic method.

For the spectral resolution, we usually use a constant step size of 1 cm$^{-1}$ in wavenumber space. This is equivalent to the one used by \citet{Line2015ApJ...807..183L} who studied similar objects.

\subsection{Radius--distance Relation}
\label{sec:radius_distance_relation}

To relate the outgoing fluxes $F_\nu^+$ computed via Eq. \eqref{eq:flux_model} to the actual ones measured by the observer ($F_\nu$), the radius and distance of the brown dwarf need to be taken into account. We therefore scale the top-of-the atmosphere fluxes $F_\nu^+$ by the usual radius--distance relation
\begin{equation}
  F_\nu = F_\nu^+ f \left(\frac{R_*}{d}\right)^2 \ ,
\end{equation}
where $d$ is the distance to the brown dwarf, $R_*$ is its radius, and $f$ is a calibration factor.

For the retrievals performed in Sect. \ref{sec:brown_dwarf_retrieval}, we choose a radius of $R_* = 1 R_\mathrm{J}$ and use distances from the \textit{Gaia} measurements. The calibration factor $f$ is treated as a free parameter that describes the uncertainties in the flux calibration of the measured spectra, but also includes the deviations of the actual brown dwarf radius from our assumed value of 1 $R_\mathrm{J}$. Lastly, $f$ also partially captures inadequacies of the forward model to describe the atmosphere of a brown dwarf in all its details, including the effect of a reduced emitting surface due to a potentially heterogeneous atmosphere.

Assuming that $f$ only includes deviations with respect to the assumed a priori radius, it can be transformed into a derived radius (in the same units as $R_*$) via
\begin{equation}
R = \sqrt{f} \ .
\label{eq:derived_distance}
\end{equation}
It should, however, be noted that in full generality, Eq.~\eqref{eq:derived_distance} does not provide a good radius estimate for the brown dwarf because $f$ usually also includes other sources of uncertainties, as described above.

\subsection{Opacities and Spectral Resolution}

This work is focused on the wavelength range of the SpeX instrument, which extends from about 0.9 $\mu$m to roughly 2.4 $\mu$m. Within this range, we account for the major absorbers that are expected to be present \citep{Line2015ApJ...807..183L}: \ce{CO2}, \ce{CO}, \ce{CH4}, \ce{H2O}, \ce{H2S}, \ce{NH3}, \ce{H2}, \ce{He}, as well as the alkali metals \ce{Na} and \ce{K}.

Calculations of the molecular line absorption cross-sections are done with our opacity calculator \texttt{Helios-k} \citep{Grimm2015ApJ...808..182G}. We use the ExoMol line lists where available \citep{Barber2006_10.1111/j.1365-2966.2006.10184.x, Yurchenko2011MNRAS.413.1828Y, Yurchenko2014MNRAS.440.1649Y, Azzam2016MNRAS.460.4063A}, and the ones provided by HITEMP \citep{Rothman2010JQSRT.111.2139R}, otherwise. The line wings are modeled by Voigt profiles \citep{Voigt1912} that describe the effects of thermal and pressure broadening. Additional details on the calculations can be found in S.~L. Grimm et al. (2020 in preparation).

Collision-induced absorption of \ce{H2}--\ce{H2} \citep{Abel_doi:10.1021/jp109441f} and \ce{H2}-He \citep{Abel2012JChPh.136d4319A} pairs are taken into account by the corresponding data provided within the HITRAN database \citep{Karman2019Icar..328..160K}.

\subsubsection{Alkali Line Absorption Cross-sections}

The treatment of the absorption cross-sections of the alkali metals Na and K is slightly different. The resonance lines of these metals are known to deviate from the usual Voigt profiles to a large degree. In particular, their far-wing line profiles are known to posses a strong non-Lorentzian behavior due to collisions of the metals with \ce{H2} molecules. Various approximations have been developed in the past to account for this behavior \citep[e.g.][]{Tsuji1999ApJ...520L.119T, burrows00, burrows03, Allard2012A&A...543A.159A, Allard2016A&A...589A..21A}.

For \texttt{Helios-r2}, we use the descriptions of the \ce{K} resonance line wings published in \citet{Allard2016A&A...589A..21A} and an updated version of the Na profiles \citep{Allard2019A&A...628A.120A}.

All other absorption lines of \ce{Na} and \ce{K} are computed based on the Kurucz line lists \citep{1995KurCD..23.....K}, using the natural line width as well as thermal and van-der-Waals broadening to describe the Voigt line profiles.

\subsection{Chemistry}

Within our forward model, we employ two different approximations for the description of the atmosphere's chemical composition. We either perform a free retrieval of the molecules' mixing ratios or use an equilibrium chemistry code to self-consistently calculate the molecular abundances.

For the equilibrium chemistry, we employ the \texttt{FastChem} model \citep{Stock2018MNRAS.479..865S}. More specifically, we here use version 2.0 of the model that features several enhancements over the previous version. Compared to \texttt{FastChem} 1.0, the new version does not require a pressure iteration and is valid for arbitrary elemental compositions (J.~W. Stock, D. Kitzmann, A.~B.~C. Patzer 2020, in preparation).

We use our standard set of about 550 chemical species that is included in the released version of \texttt{FastChem}. Due to the low temperatures expected for the brown dwarfs we aim to investigate, ions are, however, removed from the chemical network of \texttt{FastChem}. Their small abundances at low temperatures would increase the computational time of the chemistry by a factor of roughly two or more but, on the other hand, do not significantly change the number densities of the more important molecules.

A full chemistry calculation for a given temperature and pressure usually takes of the order of a few milliseconds down to one millisecond or less (in case of higher temperatures) on a single CPU. It is, thus, possible to run the chemistry model directly within the forward model.

The current version of \texttt{FastChem} does, however, not treat condensation. In equilibrium condensation chemistry models, the alkali metals are expected to condense into \ce{Na2S} and \ce{KCl} for cool T-dwarf atmospheres \citep{Marley2013cctp.book..367M}. We simulate this effect by removing K and Na from the gas phase in the upper atmosphere once the temperature--pressure profile drops below 800 K.

\subsection{Temperature Profile}

One of the most important quantities that is required for the forward model is the temperature profile. In the past, several different approaches to this problem have been used. This includes using a profile described by a nine-parameter fitting function \citep{Madhusudhan2009ApJ...707...24M}, using an approximate solution of the radiative transfer equation under the condition of radiative equilibrium in a gray atmosphere \citep[e.g.][]{Lavie2017AJ....154...91L}, or using a free layer-by-layer temperature retrieval \citep{Irwin2008JQSRT.109.1136I, Line2015ApJ...807..183L}. In this study we will explore two different scenarios, one using the gray atmosphere approximation and a modified, free-temperature retrieval based on finite elements.

The temperature profile based on the solution of the radiative transfer equation in a gray, semi-infinite atmosphere under the condition of radiative equilibrium - usually referred to as Milne's problem - is given by \citep{Mihalas1978stat.book.....M}
\begin{equation}
  T^4(\tau_R) = \frac{3}{4} \bar{T}_\mathrm{eff}^4\left[ \tau_R + q_\mathrm{H}(\tau_R) \right] \ ,
\end{equation}
where $\tau_R$ is the Rosseland optical depth, $\bar{T}_\mathrm{eff}$ an effective temperature, and $q_\mathrm{H}(\tau_R)$ the so-called Hopf function. If one assumes the Eddington approximation, i.e., the second ($K$) and the zeroth moment ($J$) of the specific intensity are related via $ 3 K = J$, then $q_\mathrm{H} = 2/3$. However, while the Eddington approximation is a good description for the deep interior of the atmosphere, its validity breaks down in the upper parts of the atmosphere, where the Eddington factor approaches unity.
We, therefore, use the exact solution of the function $q_\mathrm{H}(\tau_R)$ provided in \citet{Mihalas1978stat.book.....M}, p. 72, to describe the temperature profile.

We note that $\bar{T}_\mathrm{eff}$ should not be interpreted as the actual effective temperature of the brown dwarf's atmosphere that we aim to retrieve. It is defined as the effective temperature of a gray atmosphere, where the spectral distribution of the radiation is a pure blackbody. The atmosphere of a brown dwarf, however, is far from being gray, and its spectrum does not resemble a blackbody curve at all because it is dominated by deep molecular absorption bands. Accordingly, we determine the actual effective temperature by a post-processing procedure (see Sect.~\ref{sec:effective_temp} for details).

\subsubsection{Finite Element Approach}
\label{sec:temp_finite_elements}

\begin{figure}[th]
	\includegraphics[width=\columnwidth]{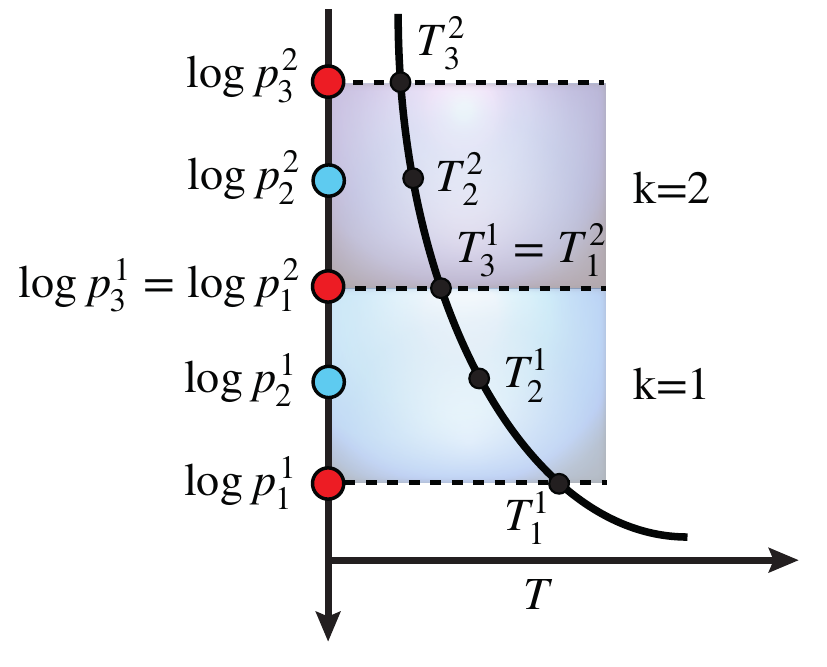}\vspace{-0.3cm}
	\caption{Schematic for approximating the temperature profile by a finite element approach. Shown are two adjacent second-order elements. The element boundaries are marked by the horizontal dashed lines, the degrees of freedom (dof) inside each element by the circles. Red circles denote degrees of freedom located at the boundaries, which are shared with the adjacent elements, blue circles refer to those inside an element. The corresponding temperatures at the dof are marked by the black dots on the temperature profile.}
	\label{fig:finite_elements}
\end{figure}

For this approach, we partition the atmosphere into a number $K_\mathrm{e}$ of non-overlapping elements, distributed in log-pressure space (see Figure~\ref{fig:finite_elements}). These elements are not aligned to and in fact are fully independent of the previously mentioned discretization of the atmosphere into layers/levels.

The discrete approximation $T_h^k(\log p)$ of the real (usually unknown) temperature $T^k(\log p)$ within each element is expressed such that
\begin{equation}
T_h^k(\log p) := \sum_{i=1}^{N_p} T_h^k(\log p_i^k) \,\ell_i^k(\log p)
\label{eq:temp_finite_local}
\end{equation}
is a polynomial of order $q$ on each element. Here, the functions $\ell_i^k$ are
the Lagrange polynomials \citep{Waring1779RSPT...69...59W, Lagrange1795} through the grid points $\log p_i^k$, i.e.,
\begin{equation}
\ell_i^k(\log p) := \prod_{\substack{j=1\\j\neq i}}^{N_p} \frac{(x-x_j)}{(x_i-x_j)} \ .
\label{eq:temp_finite_lagrange}
\end{equation}
In a finite element approach, these would be the so-called trial functions. For a given order $q$, the number of local grid points $N_p$ is given by
\begin{equation}
N_p = q + 1 \ .
\end{equation}

Figure \ref{fig:finite_elements} shows an example of two second-order elements. In that case, every element has three nodes $\log p_i^k$, associated with three degrees of freedom (dof), which are the temperatures $T_h^k(\log p_i^k) = T_i^k$. We use the continuous formulation of the approximate solution in the following, i.e., the temperatures are continuous across element boundaries. This enforces a continuous temperature profile, which is expected in a brown dwarf atmosphere, and also reduces the overall number of dof.

The temperature at the element interfaces can, in principle, also be made disjoint, allowing the solution to have discontinuous jumps from one element to the next. In terms of finite element methods, this would lead to the discontinuous Galerkin method \citep[e.g.][]{Kitzmann2016A&A...595A..90K}. This allows for more flexibility in the temperature profile -- especially in cases where the temperature profile is under-resolved -- but might also lead to unphysical discontinuities in the retrieved temperature profile. In this work, we therefore focus on the continuous version.

In the following, we assume a polynomial order larger than zero ($q > 0$). The case of $q=0$ refers to an atmosphere that is piecewise isothermal, which clearly does not satisfy the expected, continuous temperature profile.

Given the representation of the temperature in each element by Eq.~\ref{eq:temp_finite_local}, the global, piecewise continuous solution can then be written as the direct sum of all $K_\mathrm{e}$ solutions
\begin{equation}
T(\log p) \backsimeq T_h(\log p) = \bigoplus_{k=1}^{K_\mathrm{e}} T_h^k(\log p) \ .
\label{eq:temp_finite_global}
\end{equation}
This representation allows the evaluation of the temperature at any desired pressure $p$ within the atmosphere.

The total number of free parameters $N_\mathrm{T}$ for a temperature profile with $K_\mathrm{e}$ elements of order $q$ is given by
\begin{equation}
  N_\mathrm{T} = K_\mathrm{e} N_p - \left(K_\mathrm{e} - 1\right) = K_\mathrm{e} \, q + 1  \ .
  \label{eq:nb_temp_parameter}
\end{equation}

Instead of performing the calculations in Equations \eqref{eq:temp_finite_local} and \eqref{eq:temp_finite_lagrange} in the $\log p$ space, all evaluations are done on a reference element, stretching from 0 to 1 and then transformed back into the pressure space. On the reference element, the points $N_p$ are distributed according to a Gau{\ss}--Lobatto quadrature scheme \citep{Lobatto1852}.

It should be noted that the approximation of $T(\log p)$ by Eq. \eqref{eq:temp_finite_global} does not assume any specific mathematical form of the global temperature profile. Any sufficiently continuous and smooth function can be approximated by a piecewise polynomial description. Very complex temperature profiles (e.g., temperature inversions, strong gradients) may require either a high-order polynomial or a larger number of elements. While \texttt{Helios-r2} is designed to use any polynomial degree larger than unity, we normally restrict the retrieval to second-order elements. High-order polynomials are prone to suffer from so-called Runge's phenomenon \citep{Runge1901}, i.e., they are susceptible to unphysical, local oscillations.

It would be natural to use the nodal values of the temperatures $T_h^k(\log p_i^k)$ as free parameters for the retrieval. In many cases, though, this would give the temperature profile too much freedom, sometimes yielding unphysical temperature inversions in the lower atmosphere. Since one would not expect such inversions to occur based on the theory from brown dwarf atmospheres, we use a slightly different approach for the retrieval here.

The only, actually retrieved temperature is the one at the bottom of the modeled atmosphere, represented by $T_1^1$ in Figure \ref{fig:finite_elements}. For all other temperatures, we retrieve a parameter $b$, such that, e.g., $T_2^1 = b_2 T_1^1$. For first-order elements, $b$ can be interpreted as the slope between two adjacent temperature nodes. By choosing a value for $b$ of smaller or equal to unity, the temperature profile will be strictly monotonic. We found this approach to be much more stable than retrieving all temperatures individually. The same formulation can also be adapted to atmospheres with temperature inversions, by allowing $b$ to exceed a value of unity.

An example for a \texttt{Helios-r2} retrieval of an exoplanet atmosphere with a temperature inversion can be found in a separate publication \citep{Bourrier2019arXiv190903010B}. Due to the higher complexity of the temperature profile, four second-order elements are used in \citet{Bourrier2019arXiv190903010B}, while the $b$ values for the description of the profile are allowed to exceed unity.

\subsection{Calculation of Effective Temperatures}
\label{sec:effective_temp}

Since neither of the two approaches we implemented in \texttt{Helios-r} to describe the temperature profile directly yield the effective temperature $T_\mathrm{eff}$ of the brown dwarf, we estimate this parameter in a post-processing procedure. Following \citet{Line2015ApJ...807..183L}, we calculate a high-resolution spectrum for every posterior parameter combination in the wavelength range from 1 $\mu$m to 20 $\mu$m. These spectra are then integrated over these wavelengths for the total outgoing flux, which is then converted into an effective temperature by using the Stefan-Boltzmann law \citep{Stefan1879, Boltzmann1884AnP...258..291B}.

\subsection{Instrument Profile}
\label{sec:instrument_profile}

The new version of \texttt{Helios-r} can also take an instrument profile function into account when calculating theoretical spectra. This function describes the spread of the flux at a certain wavelength across several pixels on the detector due to, for example, the finite slit width of the spectrograph. We here assume this profile to have a Gaussian (normal) distribution. The calculated high-resolution spectrum from the forward model is convolved with the instrument profile $p$ to simulate the flux at a given wavenumber $\nu$ measured by each pixel on the detector
\begin{equation}
  F_{\nu,\mathrm d} = \int_{0}^{\infty} F_x \, p(x-\nu, \sigma) \mathrm d x \ ,
\end{equation}
where $\sigma$ is the standard deviation of the profile that is a characteristic of the employed instrument.
In practice, we do not perform the convolution over the entire wavenumber range, as this would be computationally extremely costly. Instead, the integration is stopped at a distance of $\lvert x - \nu \rvert = 5 \, \sigma$. Flux values outside of this range have a negligible impact on $F_{\nu,\mathrm d}$ and, thus, can be safely neglected.

\subsection{GPU Parallelization}

To make the model computationally as fast as possible, the numerically heavy part of the calculations is done on a GPU by using NVIDIA's CUDA language. This includes, for example, the interpolation of opacities, calculation of the optical depths, the solution of the radiative transfer equation, the convolution with an instrument profile, the binning of the high-resolution spectrum to the observational data, and the evaluation of the likelihood function.

The computationally most expensive part is usually the interpolation of opacities to the corresponding atmospheric temperatures and pressures. This operation, however, can be parallelized quite straightforwardly and runs much faster on a GPU than on a CPU, owing to the several thousands of cores available for simultaneous calculations on a graphics card.

The only major part of the model still running on a CPU is \texttt{FastChem}, as it cannot be efficiently parallelized for a GPU. Instead, several instances of \texttt{FastChem} are running in parallel on the CPU using OpenMP, depending on the number of available CPU cores.

In total, a typical evaluation of the forward model with 70 layers, 7100 wavelength points (corresponding to wavenumber steps of 1 cm$^{-1}$), and the equilibrium chemistry on a GeForce 2080 Ti requires about 20 ms of calculation time. Increasing the resolution to 0.3 cm$^{-1}$ (about 25,000 wavelength points) results in a calculation time of roughly 26.5 ms per model.

In principle, \texttt{Helios-r2} can also be purely run on a CPU. The calculation times, however, can then be a factor of more than 100 times higher than those obtained employing a GPU.

\section{Nested Sampling}
\label{sect:nested_sampling}

Like in the original version of \texttt{Helios-r} \citep{Lavie2017AJ....154...91L}, we also use a nested sampling approach for the Bayesian inference. Nested sampling provides an efficient way to calculate the Bayesian evidence and posterior distributions of retrieval parameters. For a full theoretical description of the nested sampling method, we refer the reader to \citet{Skilling2004AIPC..735..395S}, \citet{Feroz2008MNRAS.384..449F}, or \citet{Benneke2013ApJ...778..153B}. In the following, we provide a brief description.

\subsection{Atmospheric Retrieval in a Bayesian Framework}
\label{sec:bayesian_retrieval}

Atmospheric retrieval tries to connect observational data $\mathbf D$ of an object with a probability distribution of a parameter set $\mathbf \Theta$, that are usually connected to physically relevant properties of the observed atmosphere.
The data vector $\mathbf D$ is usually composed of a set of data points $D_j$ taken at, e.g., different wavelengths. In addition to $\mathbf D$, the observational data is also characterized  at each spectral point by a corresponding error $\sigma_j$.

Let $\mathcal M_i$ be a model with a parameter vector $\mathbf\Theta = \left\{\Theta_1, \Theta_2, ..., \Theta_N \right\}$, containing $N$ parameters $\Theta_n$. In terms of atmospheric retrieval, $\mathcal M_i$ is usually a simplified atmosphere model (\textit{forward model}) that calculates a theoretical spectrum of the observed object based on a set of input parameters, such as molecular abundances, surface gravity, or the temperature profile.
The joint prior probability distribution for $\mathbf \Theta$, $p\left( \mathbf \Theta | M_i \right)$, describes the a priori knowledge or constraints we impose on the initial distribution of the parameters $\Theta_n$.

The posterior probability distribution of $\mathbf \Theta$ for a specific model $\mathcal M_i$ applied to the data $\mathbf D$ can then be expressed following Bayes' theorem \citep{Bayes1763RSPT...53..370B}
\begin{equation}
  p\left(\mathbf \Theta | \mathbf D, \mathcal M_i \right) = \frac{p\left( \mathbf \Theta | M_i \right) \mathcal L\left(\mathbf \Theta | \mathbf D, \mathcal M_i \right)}{\mathcal Z\left(\mathbf D | \mathcal M_i \right)} \ ,
\end{equation}
where $\mathcal L\left(\mathbf \Theta | \mathbf D, \mathcal M_i \right)$ is the \textit{likelihood} and $\mathcal Z\left(\mathbf D | \mathcal M_i \right)$ the so-called \textit{Bayesian evidence}.

The likelihood function $\mathcal L\left(\mathbf \Theta | \mathbf D, \mathcal M_i \right)$ describes the probability of the model $\mathcal M_i$ to match the data $\mathbf D$, given a set of parameters $\mathbf \Theta$.
We here use the same likelihood function previously employed by, e.g., \citet{Benneke2013ApJ...778..153B}, \citet{Line2015ApJ...807..183L}, or \citet{Lavie2017AJ....154...91L}. Assuming that the observational points $j$ each possess individual Gaussian errors, the (logarithm) of $\mathcal L$ can be expressed by
\begin{equation}
\ln \mathcal L = \sum_{j=1}^{J}\left( -\frac{\left[ D_j - D_{j,\mathrm{m}}\left(\mathcal M_i, \mathbf \Theta\right) \right]^2}{2 \sigma_j^2}
- \frac{1}{2} \ln(2\pi \sigma_j^2)\right) \ ,
\label{eq:likelihood}
\end{equation}
with the theoretical observation $D_{j,\mathrm{m}}\left(\mathcal M_i, \mathbf \Theta\right)$, calculated by the forward model $M_i$ using the parameters $\mathbf \Theta$.
The Bayesian evidence is formally given by the integral
\begin{equation}
  \mathcal Z\left(\mathbf D | \mathcal M_i \right) = \int p\left( \mathbf \Theta | M_i \right) \mathcal L\left(\mathbf \Theta | \mathbf D, \mathcal M_i \right) \mathrm d \mathbf \Theta \ .
\end{equation}
Evidently, since this is a multidimensional integral over the entire parameter space, the actual, direct evaluation of this integral is quite challenging when the number of parameters is large.

The Bayesian evidence can also be used to perform model comparisons. For two different models $M_i$ and $M_j$ applied to the same data $\mathbf D$, one can compute the so-called Bayes factor
\begin{equation}
  B_{ij} = \frac{\mathcal Z\left(\mathbf D | \mathcal M_i \right)}{\mathcal Z\left(\mathbf D | \mathcal M_j \right)} \ .
\end{equation}
This factor quantifies the strength of evidence in favor of model $M_i$ over $M_j$ to represent the measured data. On the Jeffreys scale \citep{doi:10.1080/01621459.1995.10476572}, a value of 1, 3.2, and 10 correspond to no, substantial, and strong evidence in favor of model $M_i$ over $M_j$, respectively. A decisive evidence is categorized by $B_{ij} > 100$.

\subsection{Nested Sampling}

Nested sampling is essentially a method that provides the possibility of evaluating the Bayesian evidence $\mathcal Z\left(\mathbf D | \mathcal M_i \right)$ and the posterior distributions $p\left(\mathbf \Theta | \mathbf D, \mathcal M_i \right)$.  In this approach, the multidimensional integral is reduced to a one-dimensional one over the so-called \textit{prior mass}. A full description of the mathematical procedure of the nested sampling method can be found in \citet{Skilling2004AIPC..735..395S}.

As a specific implementation of the nested sampling method, we use the \texttt{MultiNest} code \citep{Feroz2008MNRAS.384..449F, Feroz2009MNRAS.398.1601F}. We use the FORTRAN version of the library and couple it directly to \texttt{Helios-r2}, written in C\texttt{++}/CUDA.

The \texttt{MultiNest} code starts by drawing $N_l$ samples from the parameter space. The points $N_l$ are referred to as live points. As mentioned in Sect. \ref{sec:bayesian_retrieval}, the parameter values are each subject to their own, individual prior distribution.
Using the \texttt{Helios-r2} forward model, a theoretical spectrum is generated for each of parameter sets of the live points. Finally, the likelihood is then evaluated via Eq.~\eqref{eq:likelihood}.

At each iteration step, \texttt{MultiNest} replaces the live point with the smallest likelihood value with a new set of values from the parameter space. This new point is chosen such that the likelihood computed for this point is higher than for the one just discarded. By repeating this process, the nested sampling will localize the regions of highest likelihood in the parameter space. To efficiently sample the parameter space, \texttt{MultiNest} employs a simultaneous
ellipsoidal nested sampling method developed by \citet{Feroz2009MNRAS.398.1601F}. We refer to \citet{Feroz2009MNRAS.398.1601F} for a detailed description on this method.

After $\mathcal Z$ is converged, the posterior distributions of the parameters $\mathbf \Theta$ are constructed by using all active live points, as well as those who have been previously removed during the iterative procedure.

The number of live points must be high enough to allow for a good coverage of the parameter space. Depending on the number of free parameters and types of priors, several hundred to thousands of live points are usually required. Ideally, convergence tests on the required number of live points should be performed to ensure that the retrieval has converged properly. For the simple test cases in Sect. \ref{sect:tests} we use between 2000 and 4000 live points, while the retrievals of the actual brown dwarf data is done with 10,000 points.

\subsection{Likelihood with Error Inflation}
\label{sect:likelhood_error_inflation}

For the retrieval of actual brown dwarf data, we use a slightly different version of the log-likelihood. Following \citet{Line2015ApJ...807..183L}, we include an inflation of the observational errors the calculation of $\ln \mathcal L$. Effectively, the usual squared error $\sigma_i^2$ is replaced by a more general data error $s_i^2$, given by
\begin{equation}
  s_j^2 = \sigma_j^2 + 10^\epsilon \ .
\end{equation}
The last factor on the right-hand side is used to slightly inflate the original observational error $\sigma_i$. With these $s_j^2$, the log-likelihood function from Eq. \eqref{eq:likelihood} is then given by
\begin{equation}
  \ln \mathcal L = -\frac{1}{2} \sum_{j=1}^{J} \frac{\left[ D_j - D_{j,\mathrm{m}}\left(\mathcal M_i, \mathbf \Theta\right) \right]^2}{s_j^2}
  - \frac{1}{2} \ln(2\pi s_j^2) \ .
\end{equation}

The exponent $\epsilon$ in the error inflation term is added as an additional retrieval parameter. For the corresponding prior of $\epsilon$, we employ the same assumptions as \citet{Line2015ApJ...807..183L} and use a uniform prior with
\begin{equation}
0.01 \cdot \min \sigma_i^2 \leq 10^\epsilon \leq 100 \cdot \max \sigma_i^2 \ .
\end{equation}

This error inflation accounts for the fact that the simplified model physics are usually not able to describe all of the details in a measured brown dwarf spectrum. Without error inflation, the nested sampling would concentrate only on the data points with the smallest errors, neglecting other important wavelength regions in the process. By inflating the error bars to a certain degree, we, thus, give the retrieval model more freedom to fit the spectrum, usually resulting in retrieved parameters that are more comparable with those expected from the theory of brown dwarf physics.

\section{Initial Testing of \texttt{Helios-r2}}
\label{sect:tests}

In this section, we first perform test retrievals on known atmospheric profiles and spectra to check that both the forward model and the retrieval are working properly. These tests are done for two different cases: a retrieval on output of the \texttt{Helios-r2} forward model itself and one on a specific model calculation from Mark Marley's \texttt{Sonora} grid of brown dwarf atmospheres (M.~S. Marley et al. 2020, in preparation).

As a specific test case, we choose an atmosphere with an effective temperature of 700 K, a $\log g$ of 4.75 in cgs units,\footnote{Unless stated otherwise, values of $\log g$ are stated in cgs units throughout this work.}, and solar elemental abundances. This roughly resembles a typical late-T dwarf. We assume a brown dwarf radius of 1 Jupiter radius, a distance of 10 pc and use an $f$ factor of 1. Tests are performed with increasing level of difficulty to evaluate the impact of each additional parameter on the posterior distributions.

\subsection{Retrieval Test on \texttt{Helios-r2} Forward Model}
\label{sec:helios_test}

For the first test, we use the \texttt{Helios-r2} forward model to produce a high-resolution spectrum. For the temperature profile, we use the model output from the \texttt{Sonora} grid for the aforementioned parameters. The high-resolution spectrum is then binned to about 150 bins from 1 to 2.4 $\mu$m.

It should be noted that due to the differences in the two atmospheric models (e.g., chemistry or opacities), the resulting effective temperature differs from the \texttt{Sonora} model atmosphere. The effective temperature derived from integrating the high-resolution spectrum of the \texttt{Helios-r2} forward model is 689 K. This value is slightly lower than the corresponding value of the original \texttt{Sonora} model (700 K).

We simulate point-wise, uncorrelated noise by shifting each point by a flux value randomly drawn from a normal distribution with a standard deviation equal to 0.2 times the median of all fluxes. The error of each bin flux is estimated by using the median fractional error of an actual SpeX observations of GJ 570 D (see Sect. \ref{sec:brown_dwarf_retrieval}).

\subsubsection{Fixed Temperature Profiles}

\begin{figure*}[t]
	\begin{center}
	  \includegraphics[width=12cm]{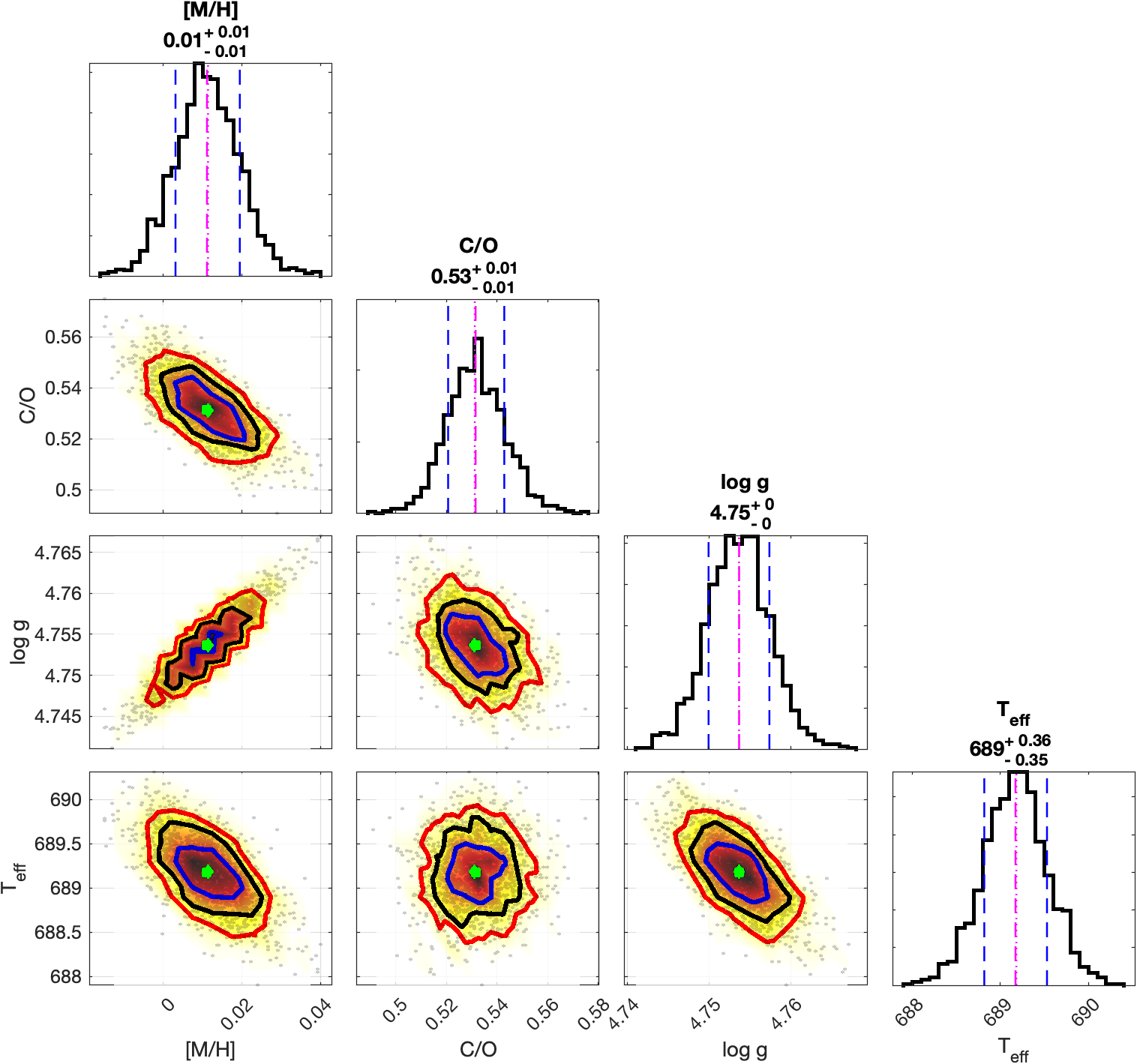}\vspace{-0.1cm}
	\end{center}
	\caption{Posterior distributions for the retrieval of a simulated \texttt{Helios-r2} spectrum using a fixed temperature profile from the \texttt{Sonora} model output. The dashed magenta-colored lines in the posterior plots refer to the location of the median value (also stated below each parameter), while the 1$\sigma$ confidence limit is denoted by the blue dashed lines. The magenta, dotted line shows the location of the best-fit model, i.e., the one with highest likelihood value. The solid blue red, and yellow lines in the two-dimensional parameter correlation plots mark the 1$\sigma$, 2$\sigma$, and 3$\sigma$ intervals, respectively. Here, the location of the median (best-fit) model is marked by green squares (diamonds). It should be noted that $T_\mathrm{eff}$ is not a directly retrieved parameter but a derived quantity.}
	\label{fig:helios_test1_post}
\end{figure*}

In the first test, we fix the temperature profile to the one from the \texttt{Sonora} output and set the $f$ parameter to its predetermined value of 1. Thus, we are only retrieving the elemental abundances, the C/O ratio, the surface gravity and, via the aforementioned post-processing procedure, the effective temperature $T_\mathrm{eff}$. This first, trivial test is purely testing the ability of the forward model to recover a mock spectrum generated by the same radiative transfer model using the same chemistry and opacities. The corresponding posteriors for these parameters are shown in Figure \ref{fig:helios_test1_post}.

Unsurprisingly, this first, simple test results in an almost perfect match of the estimated parameters to the actual values used to produce the input spectrum. All parameters are tightly constrained with quite small confidence intervals. As expected, we also find the well-known correlation between the elemental abundances and the surface gravity because of their direct influence on the atmospheric scale height.

\begin{figure*}[t]
	\begin{center}
	  \includegraphics[width=12cm]{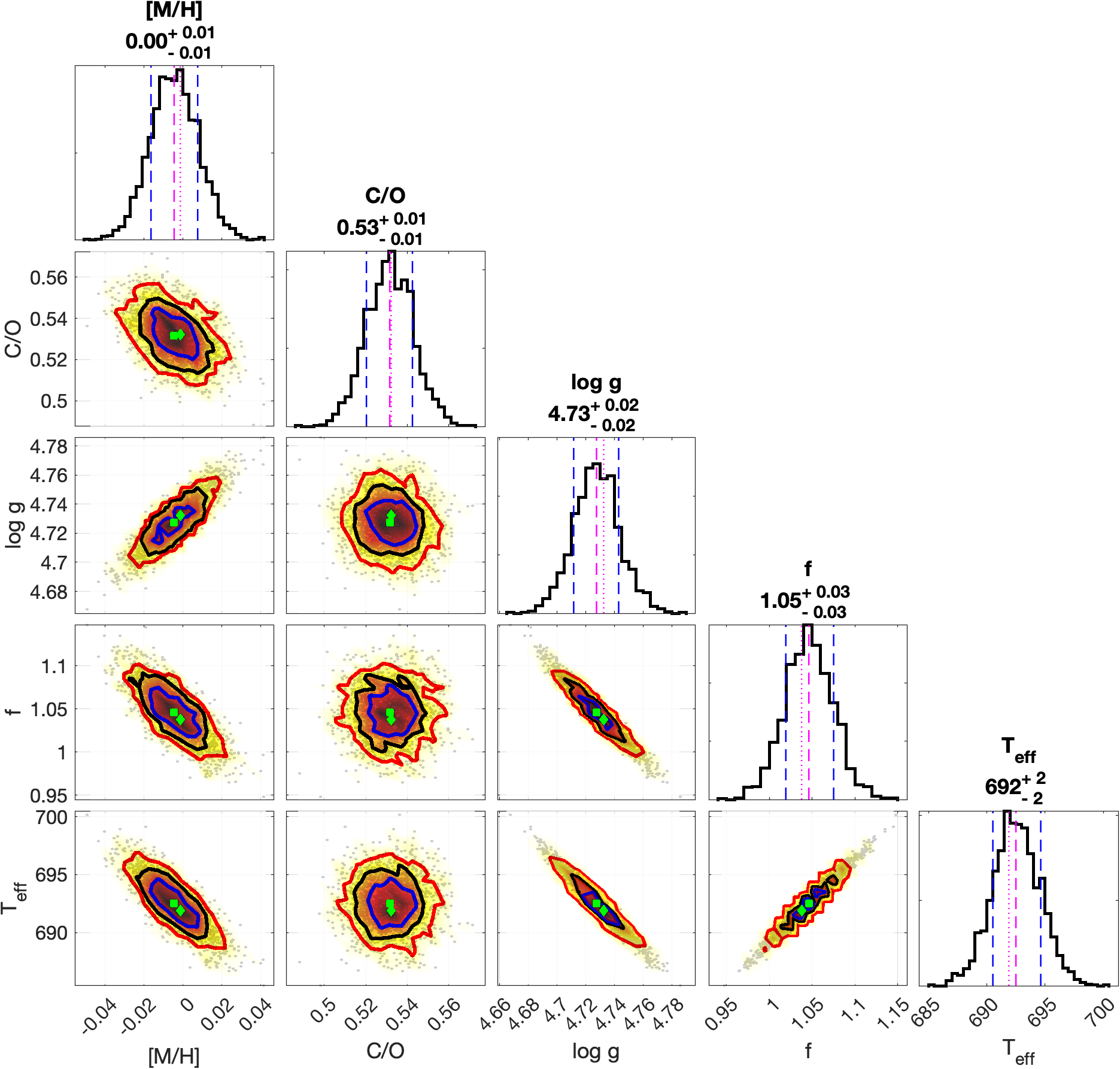}\vspace{-0.2cm}
	\end{center}
	\caption{Posterior distributions for the retrieval of a simulated \texttt{Helios-r2} spectrum using a fixed temperature profile from the \texttt{Sonora} model output. In addition to the results shown in Figure \ref{fig:helios_test1_post}, the calibration factor $f$ is added as an additional free parameter.}
	\label{fig:helios_test2_post}
\end{figure*}

In the second test, we now add the $f$ scaling as a new free parameter and repeat the retrieval. The temperature profile is still kept fixed to the one used to produce the simulated observation. The resulting posteriors are shown in Figure \ref{fig:helios_test2_post}.

Again, all parameters are well constrained and compare extremely well to their actual values. An interesting feature of this retrieval that can noticed outright, is the strong degeneracy between the effective temperature and the surface gravity. As we explain later, this will have a large impact on the actual retrieval of brown dwarf atmospheres.

\subsubsection{Free-temperature Retrieval}
\label{sec:helios_test_free_temp}

\begin{figure*}[t!]
	\vspace{0.2cm}\includegraphics[width=\textwidth]{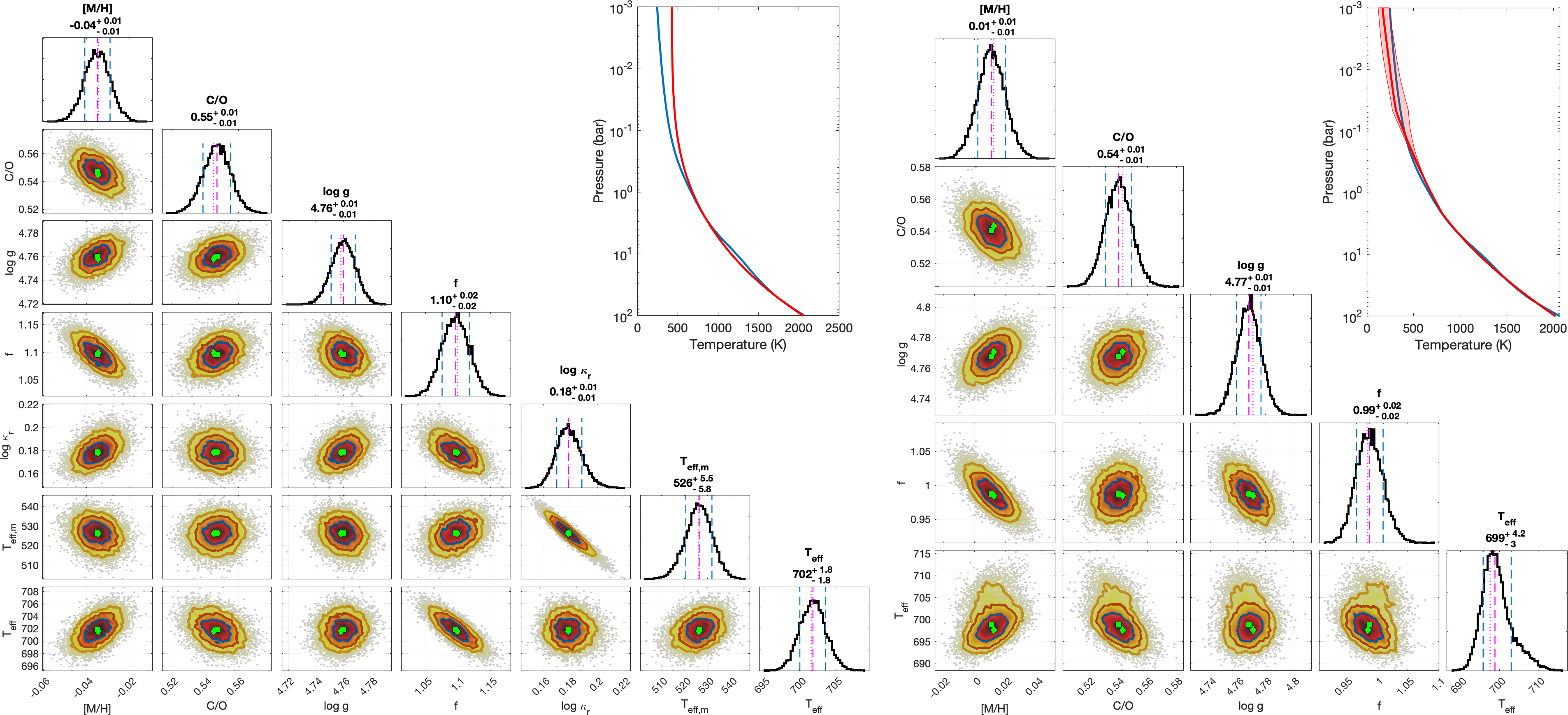}\vspace{0.2cm}
	\caption{Posterior distributions for the retrieval of a simulated \texttt{Helios-r2} spectrum with a free temperature model (right panel) and a retrieved temperature following Milne's solution (left panel). The dashed magenta-colored lines in the posterior plots refer to the location of the median value (also stated below each parameter), while the 1$\sigma$ confidence limit is denoted by the blue dashed lines. The magenta dotted line shows the location of the best-fit model, i.e., the one with highest likelihood value. The solid blue, red, and yellow lines in the two-dimensional parameter correlation plots mark the 1$\sigma$, 2$\sigma$, and 3$\sigma$ intervals, respectively. Here, the location of the median (best-fit) model is marked by green squares (diamonds). It should be noted that $T_\mathrm{eff}$ is not a directly retrieved parameter but a derived quantity. The $T_\mathrm{eff,m}$ parameter refers to the effective temperature $\bar{T}_{\mathrm{eff}}$ in Milne's solution. The panel in the upper, right corner depicts the retrieved temperature profile. The solid red line corresponds to the median profile, while the shaded, red area corresponds to the 1$\sigma$ confidence interval. The original temperature profile from the \texttt{Sonora} atmosphere model is shown in blue. Note that the free-temperature retrieval requires seven free parameters to describe the temperature profile, while Milne's solution only needs two.}
	\label{fig:helios_test3_post}
\end{figure*}

In a final test of retrieving the output of the forward model, we now also include the temperature profile in the retrieval. We study two different scenarios: a temperature profile following Milne's solution (free parameters: Rosseland opacity $\kappa_R$ and temperature $\bar{T}_{\mathrm{eff}}$), as well as a free-temperature retrieval with three second-order elements, comprised of, in total, seven free parameters ($T_1$ and six coefficients $b_i$). Details on these temperature profiles can be found in Sect. \ref{sec:effective_temp}. The posteriors of the two retrievals are shown in Figure \ref{fig:helios_test3_post}.

The left panel of Figure \ref{fig:helios_test3_post} implies that Milne's solution is only a simple approximation to the actual temperature profile. In the deep interior, the retrieved temperatures well match to the actual ones, shown in blue. This is to be expected because the approximations made for Milne's solution are valid at high optical depths (see \citet{Mihalas1978stat.book.....M} for details). On the other hand, these approximations become less valid in the upper atmosphere. As a result, Milne's solution starts to deviate from the actual profile for pressures less than 1 bar. It is, thus, not surprising that other retrieved parameters deviate from their actual values, most notably the $f$ factor that is predicted with larger value (1.10) than its actual value of 1. Apparently, the retrieval model uses this parameter to mitigate the shortcomings in the temperature profile. This emphasizes the fact that $f$ should not be seen as a purely radius-related parameter but that it also includes deviations due to assumptions made for the forward model physics.

As already mentioned in Sect. \ref{sec:effective_temp}, the free parameter $\bar{T}_{\mathrm{eff}}$ of the Milne profile should not be interpreted as the actual effective temperature. As shown in Figure \ref{fig:helios_test3_post}, the value of $\bar{T}_{\mathrm{eff}}$ is almost 200 K smaller than the effective temperature derived by integrating the high-resolution posterior spectra.

The free-temperature retrieval, on the other hand, provides an excellent representation of the temperature profile. The retrieved profile matches the original input profile in the lower atmosphere. In the upper atmosphere, the confidence intervals for the temperature profile become larger because the spectrum is almost insensitive to this part of the atmosphere at the spectral resolutions and wavelength range we are using here. The original temperature profile is, however, included in the 1-sigma envelope of the retrieved one. It should be noted, that in comparison to \citet{Line2015ApJ...807..183L}, our approach needs no additional smoothing and requires less free parameters.

\begin{figure}[h]
  \includegraphics[width=\columnwidth]{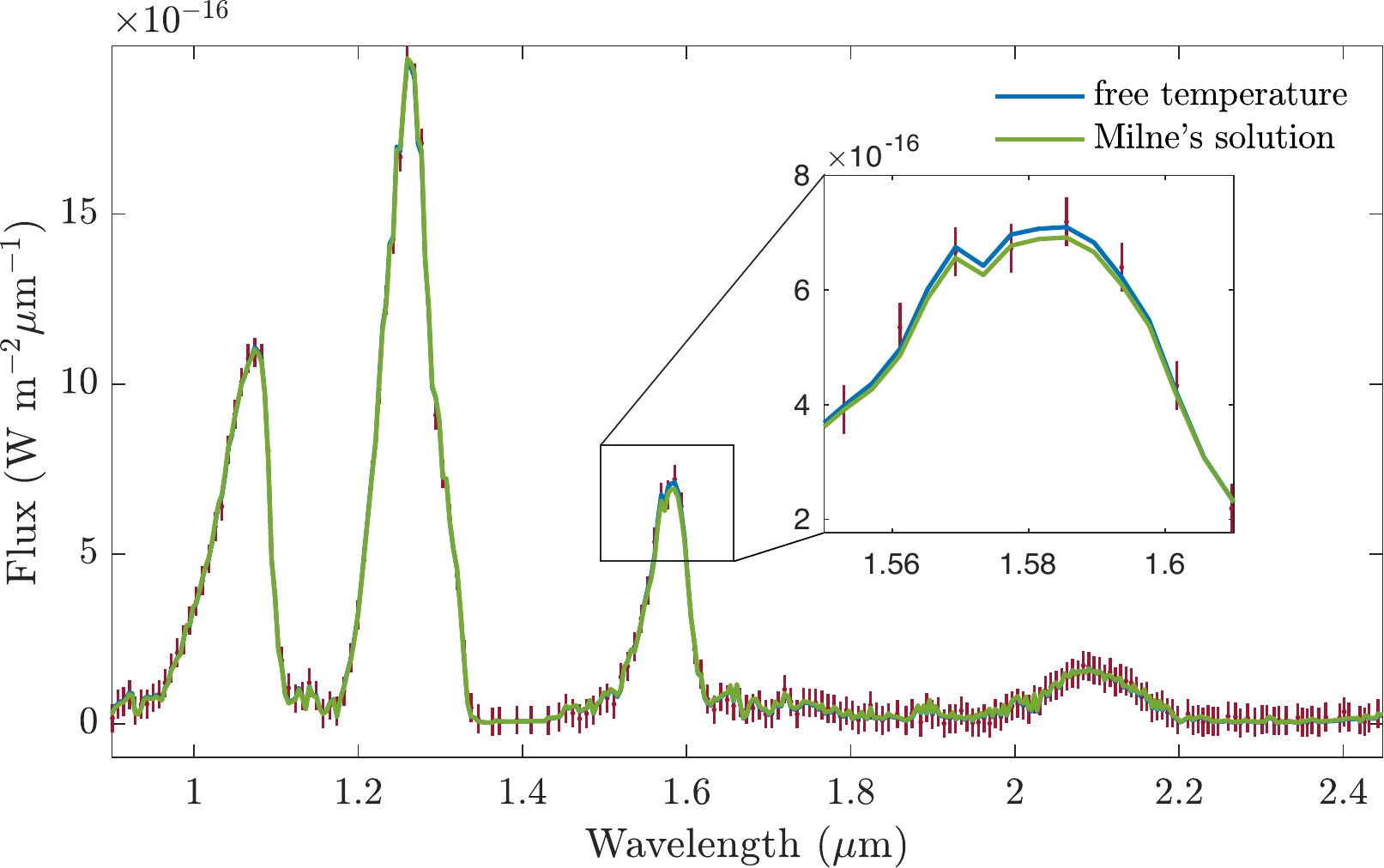}\vspace{0.2cm}
  \caption{Posterior spectra for the retrievals with the Milne solution (green) and the free-temperature model (blue). The solid lines refer to the median of all posterior spectra. The simulated observation, based on a \texttt{Sonora} model spectrum, is shown in red with its corresponding error bars. The inset plot shows a magnification of the wavelength range near 1.6 $\mu$m. Shaded areas signify the 1$\sigma$ confidence intervals of the spectra.}
  \label{fig:helios_test2_spectrum}
\end{figure}

In addition to the posterior distributions of the retrieved parameters, Figure~\ref{fig:helios_test2_spectrum} depicts the median and 1-sigma confidence levels of spectra that have been calculated for all points within the two posterior sets. As the figure suggests, there is almost no visible difference between the two distributions. Both median spectra fit the simulated data almost perfectly. Furthermore, the posteriors of the two retrievals are so tightly constrained that the 1$\sigma$ confidence ranges of the median spectra are basically invisible in Figure \ref{fig:helios_test2_spectrum}. Thus, just based on the ability of the retrieved posterior values to fit the simulated spectrum, one would not be able to exclude the Milne approximation as a valid solution of the problem.
The Bayesian evidence for the free-temperature retrieval is $\ln \mathcal Z = 14918.29$ while the one for Milne's solution is 14879.69, respectively. With a Bayes factor of $\ln B$ = 38.6, the free-temperature retrieval is decisively favored over the Milne one, even though it requires more free parameters to describe the temperature profile.
This emphasizes the decisive role played by the non-gray opacities in controlling the temperature profile.

\subsubsection{Temperature Profile Retrieval}

\begin{figure}[t!]
	\includegraphics[width=\columnwidth]{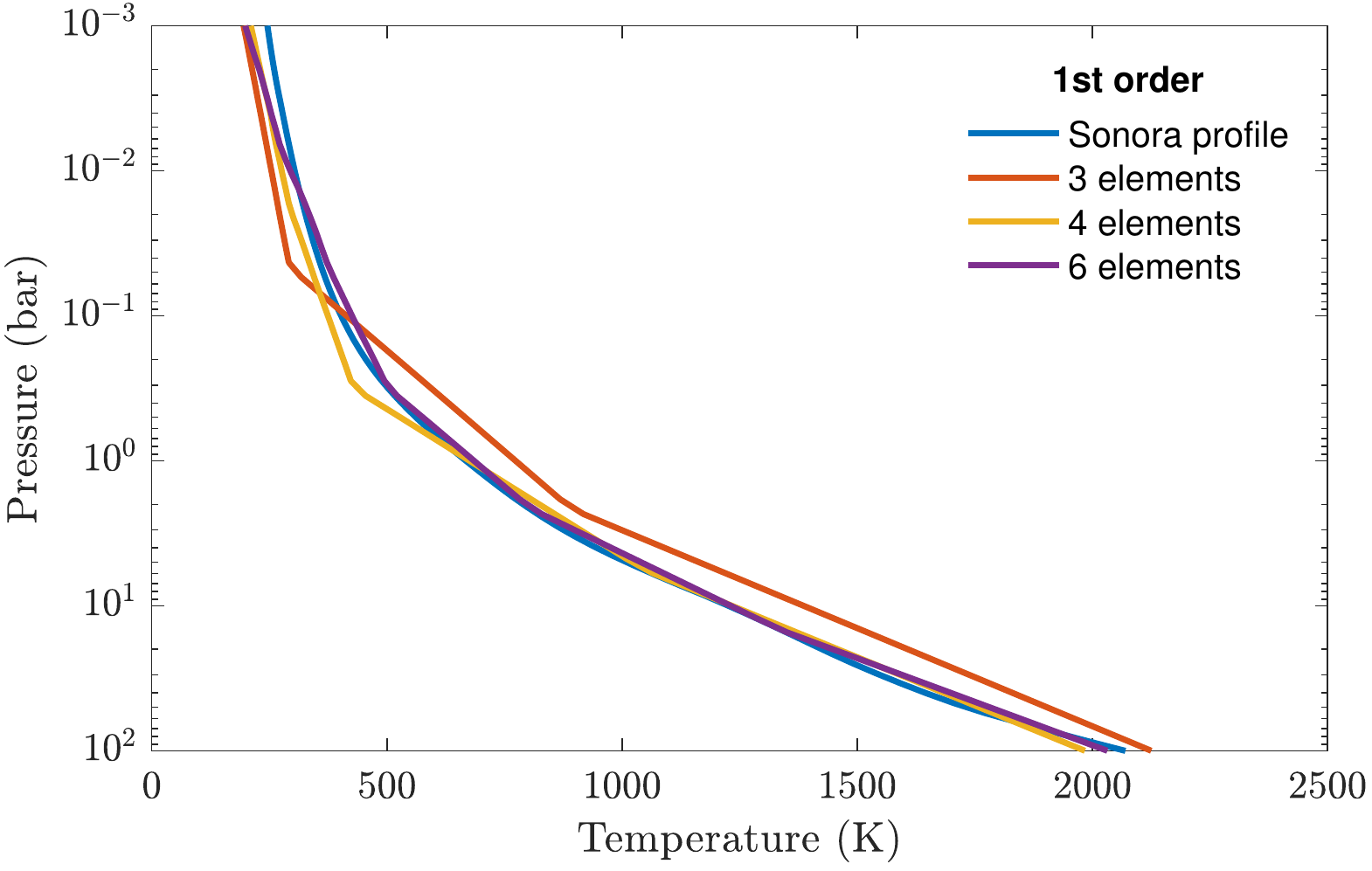}
	\includegraphics[width=\columnwidth]{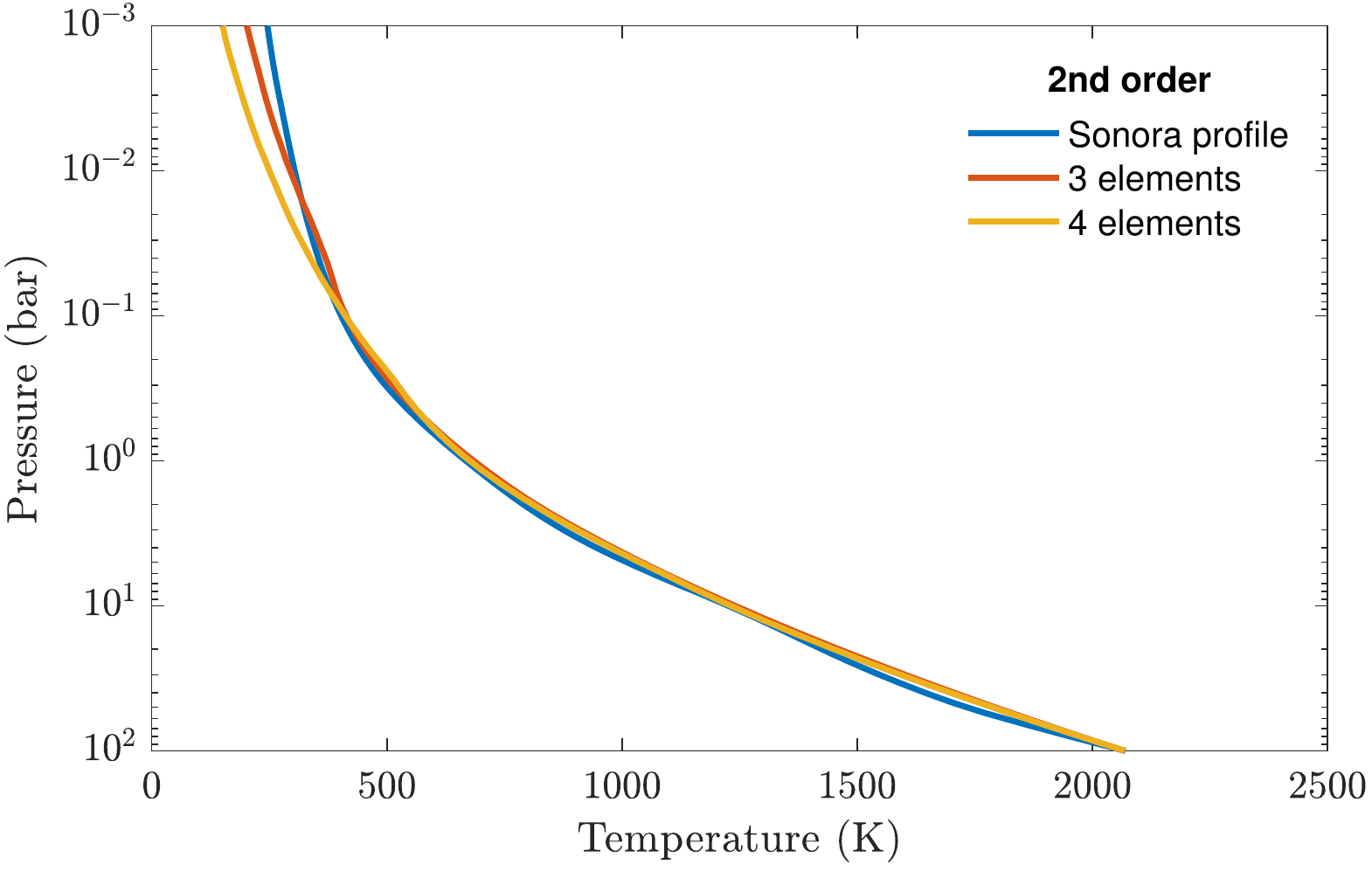}
	\caption{Impact of polynomial degree and number of elements on the retrieved, median temperature profile. The original \texttt{Sonora} profile is depicted by the blue, solid lines in both panels. Its approximation by a piecewise polynomial is shown for first-order (top panel) and second-order elements (bottom panel), for a varying total number of elements.}
	\label{fig:helios_temp_profiles}
\end{figure}

Finally, we explore the impact of the number of elements and polynomial orders on the free-temperature retrieval. As mentioned in Sect. \ref{sec:temp_finite_elements}, we normally restrict the polynomial order to a maximum of two. Figure \ref{fig:helios_temp_profiles} shows the results for first- and second-order elements.

The results clearly suggest that three first-order elements are not enough to fully describe the temperature profile. It does not provide enough degrees of freedom to cover the detailed behavior of it. On the other hand, four or six elements seem to be already sufficient. However, since they are piecewise linear functions, they still show a small level of roughness at the element boundaries.

Second-order elements are, by construction, smoother than first-order ones. Thus, they can follow the original temperature profile much more closely. This is especially  noticeable when comparing the first-order, six-element case with the second-order, three-element one. Both have the same number of degrees of freedom, but the second-order elements provide a much smoother fit. Based on the results of Section \ref{sec:temp_finite_elements}, we use either three second-order elements or six first-order ones in the actual retrievals of this study.

\begin{figure*}[t!]
	\begin{center}
		\includegraphics[width=11cm]{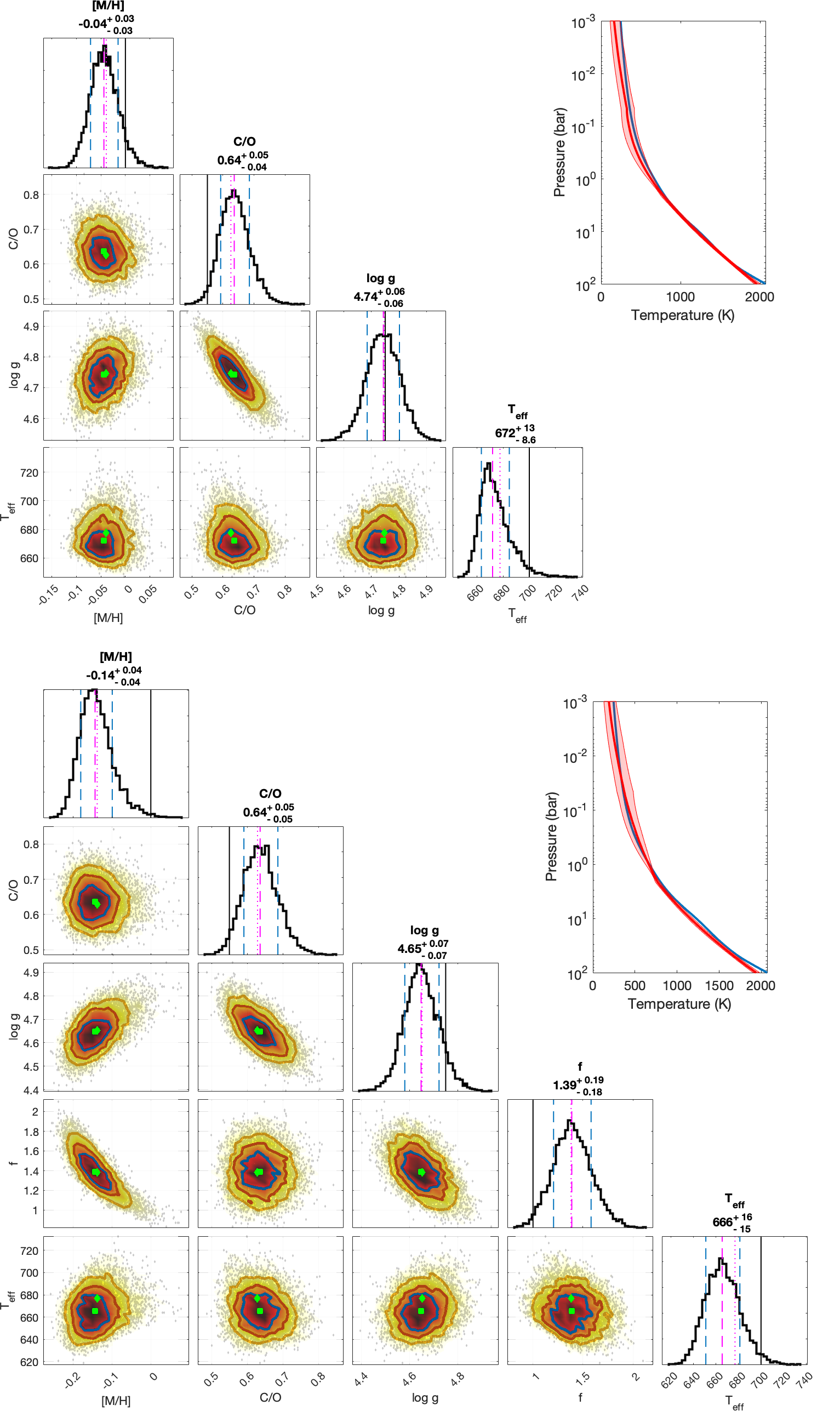}
	\end{center}
	\caption{Posterior distributions for the retrieval of a simulated observation based on a \texttt{Sonora} model spectrum (see Figure \ref{fig:helios_test3_post} for details on the posterior plots). The upper panel depicts the results for a retrieval without the $f$ calibration parameter, for the lower panel, $f$ is included as a free parameter. The original \texttt{Sonora} model parameters are marked by the solid black lines in the posterior distribution plots. The plots in the upper right corners depict the retrieved temperature profiles. The solid red lines correspond to the median profiles, while the shaded red areas indicate to the 1$\sigma$ confidence intervals. The original temperature profile from the \texttt{Sonora} atmosphere model is shown in blue.}
	\label{fig:sonora_test_post}
\end{figure*}

\subsection{Retrieval Test on \texttt{Sonora} Atmosphere}
\label{sec:sonora_comparison}

In this section, we test our retrieval model on model output from the \texttt{Sonora} grid of brown dwarf atmospheres (M.~S. Marley et al. 2020 in preparation), which uses different implementations of opacities, chemistry, and radiative transfer. As previously mentioned, we choose a model with an effective temperature of 700 K, a surface gravity of 4.75, as well as solar elemental abundances. 

The \texttt{Sonora} element abundances, however, are always given with respect to the bulk composition, i.e. they include the elements in the gas phase, as well as those present in condensates. The retrieval model, on the other hand, is only sensitive to the abundances in the gas phase. Thus, the retrieved metallicities and C/O ratios can differ from the original \texttt{Sonora} input values. In order to provide a more consistent comparison with the rainout chemistry of \texttt{Sonora}, we remove several heavier elements from the \texttt{FastChem} equilibrium chemistry, such as Fe, Mg, or Si. 

We also again assume a stellar radius of 1 Jupiter radius, a distance of 10~pc, and an $f$ factor of 1. The simulated observation is created the same way as for the \texttt{Helios-r2} test case. The comparison is performed for two different cases: with and without the calibration factor $f$. The temperature profile is freely retrieved in both cases.

The posteriors for the retrieval without the $f$ parameter, shown in the upper panel of Figure \ref{fig:sonora_test_post}, are well constrained, with only small standard deviations. The retrieved values for $\log g$ and the effective temperature are a bit less accurate than in the previous \texttt{Helios-r2} test retrieval, owing to differences in the two atmospheric models. This is most likely caused by different opacity line lists, differences in the chemical networks, or radiative transfer methods. Nonetheless, the retrieved values are quite close to the ones from the \texttt{Sonora} grid. 

The metallicity derived by \texttt{Helios-r2} is slightly sub-solar, while the C/O shows an enrichment in carbon compared to solar element abundances used by \texttt{Sonora}. This increased C/O ratio is expected because the latter model considers the removal of chemical species via condensation. Since this includes also oxygen-bearing condensates, oxygen will have a smaller than solar elemental abundance in the gas phase, which results in a super-solar C/O ratio.

In a second test, we now add the $f$ factor to the retrieval (Figure \ref{fig:sonora_test_post}, lower panel). The posteriors imply that the $f$ parameter has a very strong impact on the other retrieval parameters. Instead of the expected value of 1, we obtain the much higher value of 1.39. This clearly also influences the other posterior distributions compared to the previous case without $f$. The effective temperature decreased slightly, while the surface gravity and the metallicity are affected more strongly. The retrieval seems to be partially misled by the $f$ parameter and uses it to mitigate differences in the two atmospheric models. As already mentioned in Section \ref{sec:helios_test_free_temp}, $f$ should not be viewed as a parameter that is only related to the brown dwarf radius, as it also includes contributions due to, e.g., choices of molecular line lists or model physics assumptions.

The temperature profile is retrieved quite accurately in both cases. It follows the \texttt{Sonora} one in the lower atmosphere but is slightly shifted to lower temperatures in the case that includes the calibration factor as a free parameter. As expected, the profile has a wider confidence interval in the upper atmosphere where the spectrum becomes insensitive to the temperature. The \texttt{Sonora} profile, however, is still included within this interval in both cases.

The Bayes factor of these two retrievals is 1.37. Evaluating this factor using the Jeffreys scale (see Sect. \ref{sec:bayesian_retrieval}) implies that there is very weak to no evidence of favoring either of the two different models.

\begin{figure}[ht]
	\includegraphics[width=\columnwidth]{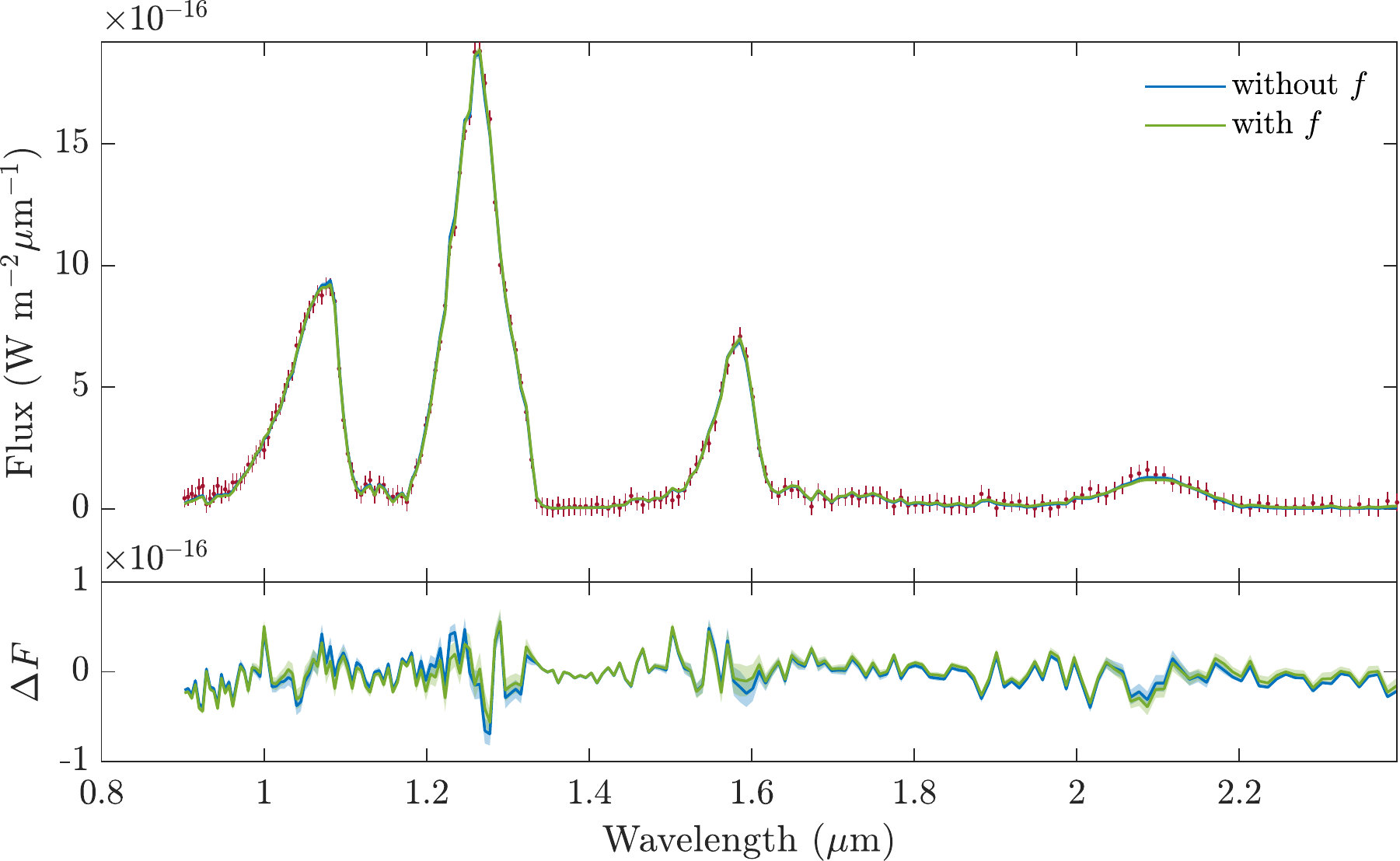}
	\caption{Posterior spectra and residuals for the retrievals of the simulated observation based on a \texttt{Sonora} model spectrum using a fixed temperature profile. The solid lines refer to the median of all posterior spectra for a retrieval including the calibration factor (green) and the one without $f$ (blue). Shaded areas signify the 1$\sigma$ confidence intervals of the spectra. The simulated observation, based on a \texttt{Sonora} model spectrum, is shown in red with its corresponding error bars. }
	\label{fig:sonora_test1_spectrum}
\end{figure}

The median spectra of both retrieval tests are shown in Figure \ref{fig:sonora_test1_spectrum}. The figure clearly implies that both retrievals worked -- in the sense that they provide a solution to fit the simulated observational data. Like in the previous case of the \texttt{Helios-r2} retrieval test, both cases -- despite their different retrieved parameters -- are almost indistinguishable. Larger differences between the two cases can be identified in the region near the 1 $\mu$m peak. Here, the retrieval including the $f$ factor seems to fit the \texttt{Sonora} model spectrum better. Thus, the $f$ parameters tries to provide a more accurate fit in this region.

After comparing the various opacity sources with those in the \texttt{Sonora} model (D. Saumon, private communication), we found the cause of these differences in the resonance ling wings of the alkali metals \ce{K} and \ce{Na}. The \texttt{Sonora} model currently uses older versions of the profiles from N. Allard that were limited to a perturber density of $10^{19}$ cm$^{-3}$. The alkali opacity of \texttt{Helios-r2}, on the other hand, is based on newer calculations \citep{Allard2016A&A...589A..21A, Allard2019A&A...628A.120A} that are now valid for H$_2$ densities up to $10^{21}$ cm$^{-3}$. Furthermore, the new versions of the alkali line wings fall off much faster at large distances from the line center than the older ones used in the \texttt{Sonora} model. Thus, even the impact of \ce{Na} can still be seen at 1.1 $\mu$m in the \texttt{Sonora} model, whereas the new sodium far-wing line profiles used in \texttt{Helios-r2} have already decayed to undetectable values.

\section{Brown Dwarf Atmospheric Retrieval}
\label{sec:brown_dwarf_retrieval}

In this section, we apply \texttt{Helios-r2} to measured spectra of three brown dwarfs: \object{GJ 570 D}, \object{$\epsilon$ Indi Ba}, and \object{$\epsilon$ Indi Bb}.

\object{GJ 570 D} has been identified as a T7.5 dwarf by \citet{Burgasser2004AJ....127.2856B} and \citet{Burgasser2006ApJ...637.1067B}. Considered a brown dwarf benchmark object, GJ 570 D has been the subject of several studies in the past. This includes retrievals using pre-calculated atmosphere grids \citep{saumon06, Geballe2001ApJ...556..373G, Saumon2012ApJ...750...74S} as well as retrieval approaches with a Markov chain Monte Carlo (MCMC) model \citep{Line2015ApJ...807..183L}.

The binary brown dwarfs in the $\epsilon$ Indi system consist of an early T1.5 dwarf ($\epsilon$ Indi Ba) and $\epsilon$ Indi Bb, a T6 dwarf \citep{Burgasser2006ApJ...637.1067B}. They have previously been studied using grids of brown dwarf models by, e.g., \citet{Kasper2009ApJ...695..788K} or \citet{King2010AA...510A..99K}. Dynamical masses of the binaries have been reported by \cite{dieterich18}.

\subsection{Observations}

\paragraph{GJ 570 D}

Our spectrum of GJ 570 D has been taken by the SpeX instrument, spanning the wavelength range from 0.8 to about 2.5 $\mu$m \citep{Burgasser2006ApJ...639.1095B}. The spectral resolution varies between about 85 and 300 throughout the spectrum. The spectrograph's slit width for this observation is 0"5. With an image scale of $0\farcs15$ per pixel, the spectral flux at a given wavelength in the spectrum is, thus, sampled onto 3.3 pixels on the CCD. This oversampling is accounted for in our retrieval by using an appropriate instrument profile (see Sect. \ref{sec:instrument_profile} for details). In order to obtain the wavelength-dependent standard deviation for the instrument profile, we estimate the local width of a pixel in wavelengths by using the values of neighboring wavelength points in the spectrum. The result multiplied by 3.3 is then approximately the FWHM of the Gaussian instrument profile.

As mentioned by \citet{Line2015ApJ...807..183L}, the oversampling also results in neighboring pixel being not statistically independent because the flux information is partly duplicated. Therefore, we follow the approach of \citet{Line2015ApJ...807..183L} and use only every third pixel in our retrieval.

We use 2MASS photometric data to flux-calibrate the spectrum of GJ 570 D. The spectrum is scaled by a multiplicative factor that is separately computed for the $J$ (15.32 $\pm$ 0.05 mag), $H$ (15.27 $\pm$ 0.09 mag), and $K_S$ (15.24 $\pm$ 0.16 mag) bandpasses \citep[see][for details]{Cushing:2005ed}. Uncertainties in the scale factor take into account spectral measurement errors and photometric uncertainties. We adopt the weighted mean of these three values for our final flux calibration scale factor. The calibrated spectrum of GJ 570 D is shown in Figure \ref{fig:gj_retrieval_spectra}.

\paragraph{Epsilon Indi Ba \& Bb}

The two brown dwarfs in the $\epsilon$ Indi system have been previously observed by \citet{King2010AA...510A..99K} using the FORS2 instrument at the Very Large Telescope (VLT) in the optical wavelength range and the ISAAC spectrograph (VLT) in the near infrared and infrared. Details on the calibration of the spectrum can be found in \citet{King2010AA...510A..99K}.
In the following, we focus on the near-infrared ISAAC measurement that covers about the same wavelength range as the SpeX measurement of GJ 570 D. With about 20,000 wavelength points, the spectral resolution of the ISAAC measurement is much higher than the one provided by the SpeX prism.

Using the full resolution of 20,000 wavelength points, however, is problematic in the framework of the nested sampling Bayesian retrieval used in this study. The definition of the likelihood function in Eq. \eqref{eq:likelihood} is essentially the $\chi^2$ distance between the measured flux values and those predicted from the forward model. At high dimensions, this distance is known to lose its mathematical meaning \citep{Beyer10.1007/3-540-49257-7_15}, which is commonly referred to as the \textit{curse of dimensionality}. It should be noted that due to its lower spectral resolution, spectra obtained with the \textit{JWST} will not have this issue.

To avoid this issue, we integrate the 20,000 wavelength bins to a lower resolution. In total, we use about 400 bins, which is equivalent to the full resolution of the SpeX instrument. Due to the binning to a resolution that is much lower than the original one, no instrument profile has to be accounted for, and this effect is, therefore, neglected for $\epsilon$ Ind Ba and Bb. The spectra of $\epsilon$ Indi Ba \& Bb are shown in Figure \ref{fig:epsind_retrieval_spectra}.

\subsection{Retrieval Parameters}

\begin{deluxetable}{lll}[t]
	\tablecaption{Summary of Retrieval Parameters and Prior Distributions Used for the Free and Equilibrium Chemistry Models.\label{tab:retrieval_parameter}}
	\tabletypesize{\scriptsize}
	\tablehead{
		\colhead{Parameter} &
		\multicolumn{2}{c}{Prior} \\
		\colhead{} &
		\colhead{Type} &
		\colhead{Values}
	}
	\startdata
	$\log g$   & uniform   & 3.5 -- 6.0 \\
	$d$   & Gaussian & measured\tablenotemark{a} \\
	$f$        & uniform   & 0.1 - 5.0 \\
	$T_1$      & uniform   & 1000 -- 3000 \\
	$b_i$      & uniform   & 0.3 -- 0.95 \\
  $\epsilon$ & uniform   & min($\sigma_j$) -- max($\sigma_j$) \\
	\hline
	\textit{Equilibrium chemistry} & & \\
	M/H  & uniform & 0.1 -- 5.0 \\
	C/O  & uniform & 0.1 -- 4.0 \\
	\hline
	\textit{Free chemistry} &  & \\
	$x_i$ & log-uniform & $10^{-12}$ -- $10^{-1}$ \\
	\enddata
	\tablenotetext{a}{We use distances inferred from \textit{Gaia} parallax measurements \citep{Gaia2018yCat.1345....0G}. For the GJ 570 system, the measured distance is $5.8819 \pm 0.0029$ pc, for $\epsilon$ Ind $3.6389 \pm 0.0033$ pc.}
\end{deluxetable}

For each of the three brown dwarfs, we split the retrieval calculations into two different categories, a first one assuming equilibrium chemistry and a second using a free chemistry approach. A summary of all retrieval parameters is given in Table \ref{tab:retrieval_parameter}.
For the equilibrium chemistry model, we use the overall metallicity M/H (assuming solar element abundance ratios) and the C/O ratio as free parameters. In the free chemistry approach, we retrieve for the mixing ratios $x_i$ of \ce{H2O}, \ce{NH3},  \ce{CH4}, \ce{CO}, \ce{CO2}, \ce{H2S}, and K. We note that in contrast to \citet{Line2015ApJ...807..183L}, we do not use the mixing ratio of sodium as a free parameter. The Na abundance is obtained from the potassium mixing ratio by using their solar elemental abundance ratio. As mentioned in Sect. \ref{sec:sonora_comparison}, by using the new far-wing resonance line-wing profiles, our calculated spectra are insensitive to sodium beyond a wavelength of 1 $\mu$m. 

The species' mixing ratios as a function of pressure are assumed to be isoprofiles, i.e., the retrieved abundances should be interpreted as the mean abundances within the visible atmosphere. The abundances of \ce{H2} and \ce{He} are obtained by assuming that they make up the rest of the atmosphere, using the solar ratio of their elemental abundances.

The metallicity and C/O ratio are obtained in a post-process procedure. The C/O ratio is computed by dividing the sum of the carbon-bearing species by the sum of oxygen-bearing species, each weighted by the number of carbon or oxygen atoms present in the molecule. The metallicity [M/H] is approximated by summing up the constant mixing ratios for each species weighted by the number of metal atoms and divided by the abundance of hydrogen. The result is then compared to the sum of solar metals relative to hydrogen. 

In addition, we introduce the distance $d$ as a retrieval parameter. We use the measured distances and the corresponding errors with a Gaussian prior (see Table \ref{tab:retrieval_parameter}). This procedure will propagate the error in the measured distances through all other retrieval parameters.

The temperature profile is assumed to be characterized by either six first-order or three second-order elements. In both cases, the profile is described by seven free parameters. Thus, for a free chemistry retrieval, we have in total 18 free parameters, while the equilibrium chemistry approach requires 13.

\begin{deluxetable*}{llccl}
	\tabletypesize{\scriptsize}
	\tablecaption{Summary of Retrieved Parameters for the Three Brown Dwarfs and Comparison with Previously Reported Values. \label{tab:retrieval_results}}
	\tablecolumns{5}
	\tablewidth{0pt}
	\tablehead{
		\colhead{} &
		\colhead{Parameter} &
		\multicolumn{2}{c}{This work} &
		\colhead{Previous work} \\
		\colhead{} &
		\colhead{} &
		\colhead{ \makecell{Equilibrium \\ Chemistry} } &
		\colhead{ \makecell{Free \\ Chemistry} } &
		\colhead{}
	}
	\startdata
	\textbf{GJ 570 D} &  $T_\mathrm{eff}$ (K) & $730^{+18}_{-17}$ & $703^{+17}_{-20}$ & $714^{+20}_{-23}$ \citep{Line2015ApJ...807..183L} \\
	& & &  & $759 \pm 63$ \citep{Filippazzo2015ApJ...810..158F}  \\
	& & &  & $780-820$ \citep{Burgasser2006ApJ...639.1095B} \\
	& & &  & $800 - 820$ \citep{saumon06} \\
	& & &  & $900$ \citep{Testi2009} \\
	& & &  & $948 \pm 53$ \citep{delBurgo2009} \\
	& $\log g$ & $4.61^{+0.08}_{-0.08}$ & $5.01^{+0.13}_{-0.19}$ & $4.5^{+0.5}_{-0.5}$ \citep{delBurgo2009} \\
	& & &  & $4.76^{+0.27}_{-0.28}$ \citep{Line2015ApJ...807..183L} \\
	& & &  & $4.90 \pm 0.5$ \citep{Filippazzo2015ApJ...810..158F} \\
	& & &  & $5.0$ \citep{Testi2009} \\
	& & &  & $5.09 - 5.23$ \citep{saumon06}\\
	& & &  & $5.1$ \citep{Burgasser2006ApJ...639.1095B} \\
	&$R$ (R$_\mathrm{J}$) & $1.00^{+0.10}_{-0.09}$ & $1.13^{+0.05}_{-0.06}$ & $1.14^{+0.10}_{-0.09}$ \citep{Line2015ApJ...807..183L} \\
	& C/O  & $0.83^{+0.09}_{-0.08}$ & $1.11^{+0.09}_{-0.09}$ & 0.95 -- 1.25\tablenotemark{a}, 0.70\tablenotemark{b} \citep{Line2015ApJ...807..183L} \\
	& [M/H] & $-0.15^{+0.05}_{-0.04}$ & $-0.13^{+0.06}_{-0.08}$ & -0.29 -- -0.04\tablenotemark{a}, -0.15\tablenotemark{b} \citep{Line2015ApJ...807..183L} \\
	& $\ln \mathcal Z$ & 4775.73 & 4775.06 & \\
	\hline
	\textbf{$\boldsymbol \epsilon$ Indi Ba} &  $T_\mathrm{eff}$ (K) & $1339^{+19}_{-19}$ & $1420^{+16}_{-16}$ & 1250 -- 1300 \citep{Kasper2009ApJ...695..788K}\\
	& & & & 1300 -- 1400 \citep{King2010AA...510A..99K}\\
	& $\log g$ & $5.49^{+0.06}_{-0.10}$ & $5.62^{+0.07}_{-0.07}$ & 5.2 -- 5.3 \citep{Kasper2009ApJ...695..788K}\\
	& & & & $5.27 \pm 0.09$ \citep{dieterich18}\tablenotemark{c}\\
	& & & & 5.50 \citep{King2010AA...510A..99K}\\
	&$R$ (R$_\mathrm{J}$) & $0.73^{+0.02}_{-0.02}$ & $0.55^{+0.01}_{-0.01}$ & \\
	& C/O  & $0.44^{+0.04}_{-0.03}$ & $0.95^{+0.02}_{-0.03}$  &\\
	& [M/H]  & $-0.70^{+0.06}_{-0.07}$ & $0.89^{+0.17}_{-0.23}$ & $-0.2$ \citep{King2010AA...510A..99K}\\
	& $M$ ($M_\mathrm{J})$ & $67^{+8.4}_{-12}$ & $50^{+7.8}_{-6.8}$ & $75\pm0.82$ \citep{dieterich18} \\
	& $\ln \mathcal Z$ & 4319.21 & 4352.01 & \\
	\hline
	\textbf{$\boldsymbol \epsilon$ Indi Bb} &  $T_\mathrm{eff}$ (K) & $768^{+26}_{-25}$ & $992^{+22}_{-21}$ &875 -- 925 \citep{Kasper2009ApJ...695..788K}\\
	& & & & 880 -- 940 \citep{King2010AA...510A..99K}\\
	& $\log g$ & $5.11^{+0.05}_{-0.05}$ & $4.85^{+0.17}_{-0.19}$ & 4.9 -- 5.1 \citep{Kasper2009ApJ...695..788K}\\
	& & & & $5.24 \pm 0.09$ \citep{dieterich18}\tablenotemark{c}\\
	& & & & 5.25 \citep{King2010AA...510A..99K}\\
	&$R$ (R$_\mathrm{J}$) & $0.73^{+0.02}_{-0.02}$ & $0.71^{+0.04}_{-0.03}$ & \\
	& C/O  & $1.21^{+0.09}_{-0.08}$  & $0.84^{+0.07}_{-0.07}$  &\\
	& [M/H]  & $-0.30^{+0.06}_{-0.06}$ & $-0.34^{+0.12}_{-0.11}$ & $-0.2$ \citep{King2010AA...510A..99K}\\
	& $M$ ($M_\mathrm{J})$ & $77^{+2.2}_{-4.2}$ & $15^{+6.0}_{-4.6}$ & $70.1\pm0.68$ \citep{dieterich18} \\
	& $\ln \mathcal Z$ & 4366.03 & 4417.71 & \\
	\enddata
	\tablenotetext{a}{Retrieved parameter}
	\tablenotetext{b}{Derived from post-process chemistry model}
	\tablenotetext{c}{Derived parameter, based on the measured dynamical mass and assuming $R = 1 \pm 0.1 R_\mathrm{J}$}
\end{deluxetable*}

A summary of the retrieval results for all three brown dwarfs is given in Table \ref{tab:retrieval_results}. Additionally, the table also lists the corresponding parameters obtained by other studies for comparison. The posterior distributions are shown and discussed within the next subsections.

\subsection{Retrieval of GJ 570 D}

\begin{figure*}[t!]
	\begin{center}
		\includegraphics[width=10cm]{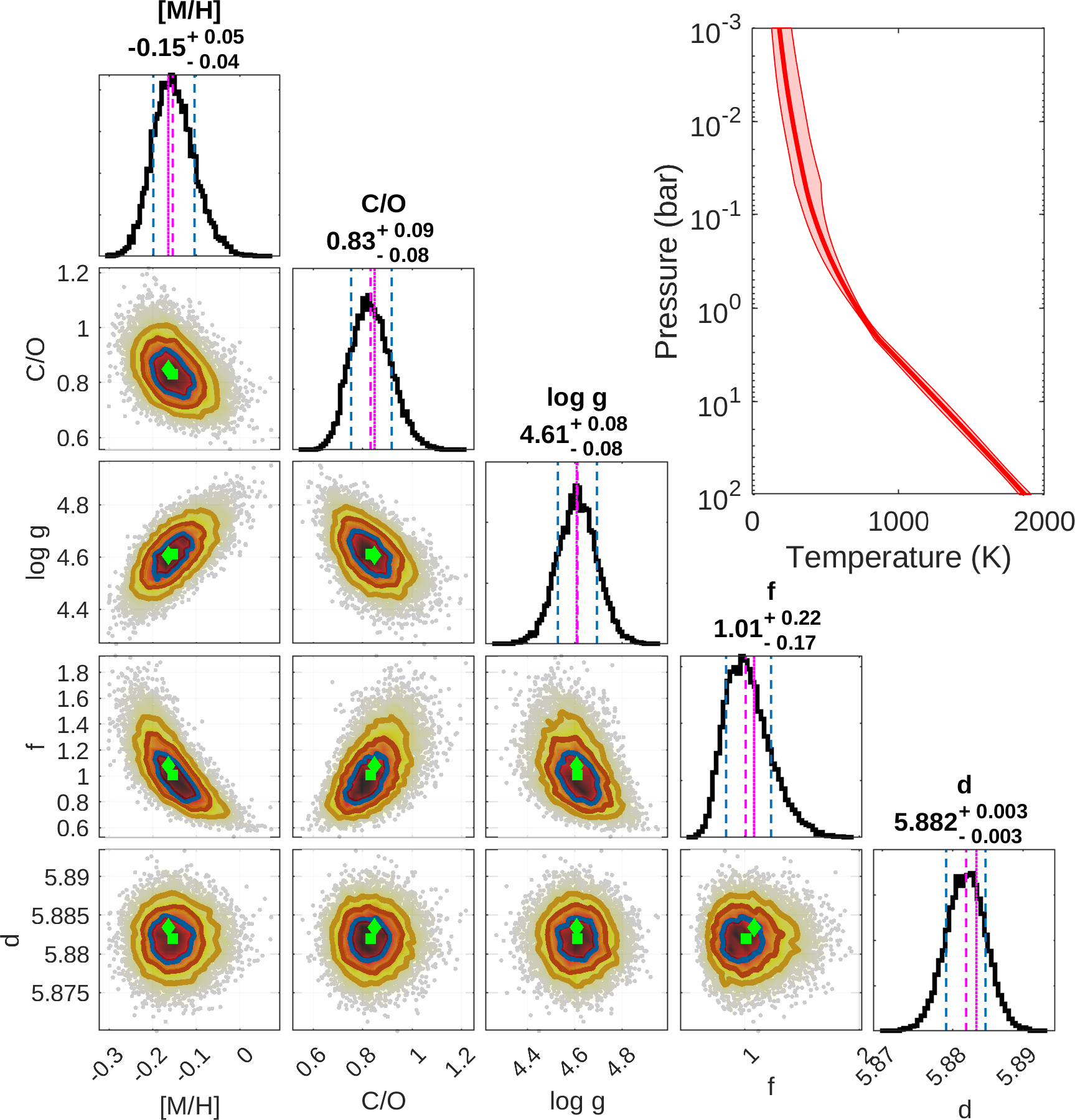}\vspace{0.2cm}
		\includegraphics[width=8cm]{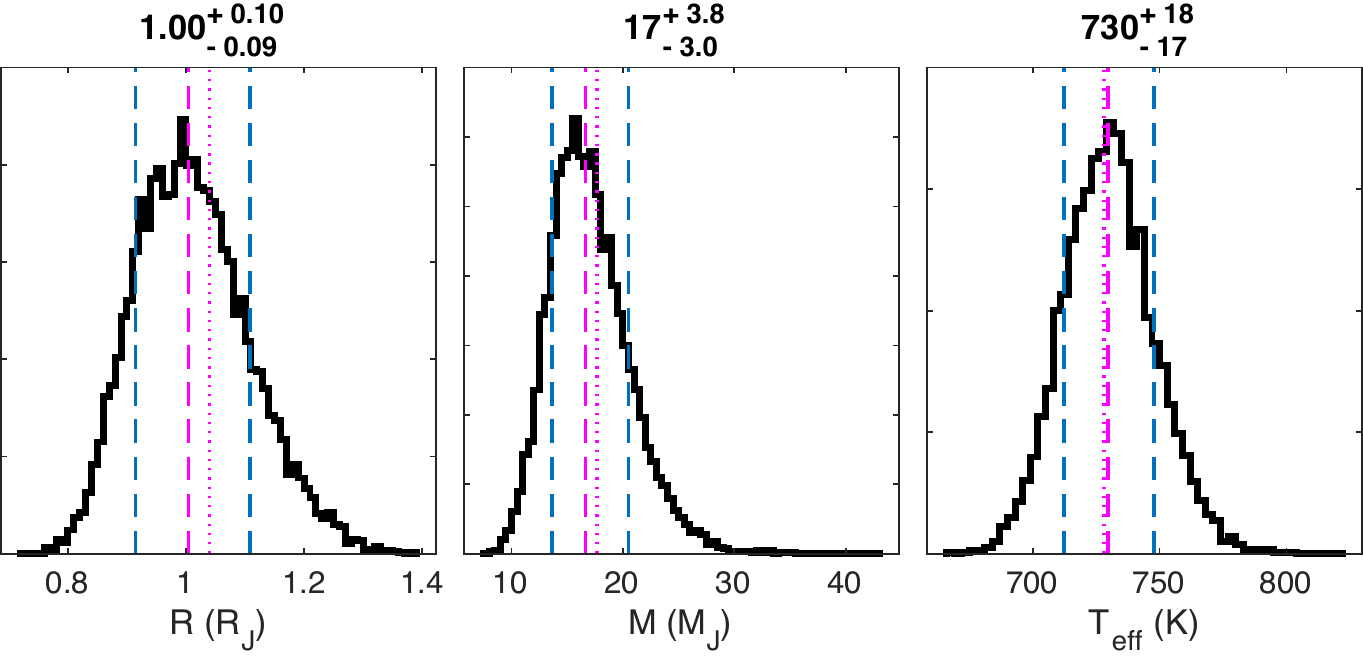}
	\end{center}
	\caption{Posterior distributions for the retrieval of GJ 570 D assuming equilibrium chemistry (see Figure \ref{fig:helios_test3_post} for details on the posterior plots). The upper panel summarizes the results for the direct retrieval parameters. The lower panel depicts posterior distributions for derived quantities. The plot in the upper right corner shows the retrieved temperature profile. The solid line corresponds to the median profile, while the shaded area indicates to the 1$\sigma$ confidence interval.}
	\label{fig:gj_retrieval_eq}
\end{figure*}

\begin{figure*}[t]
	\begin{center}
		\includegraphics[width=0.9\textwidth]{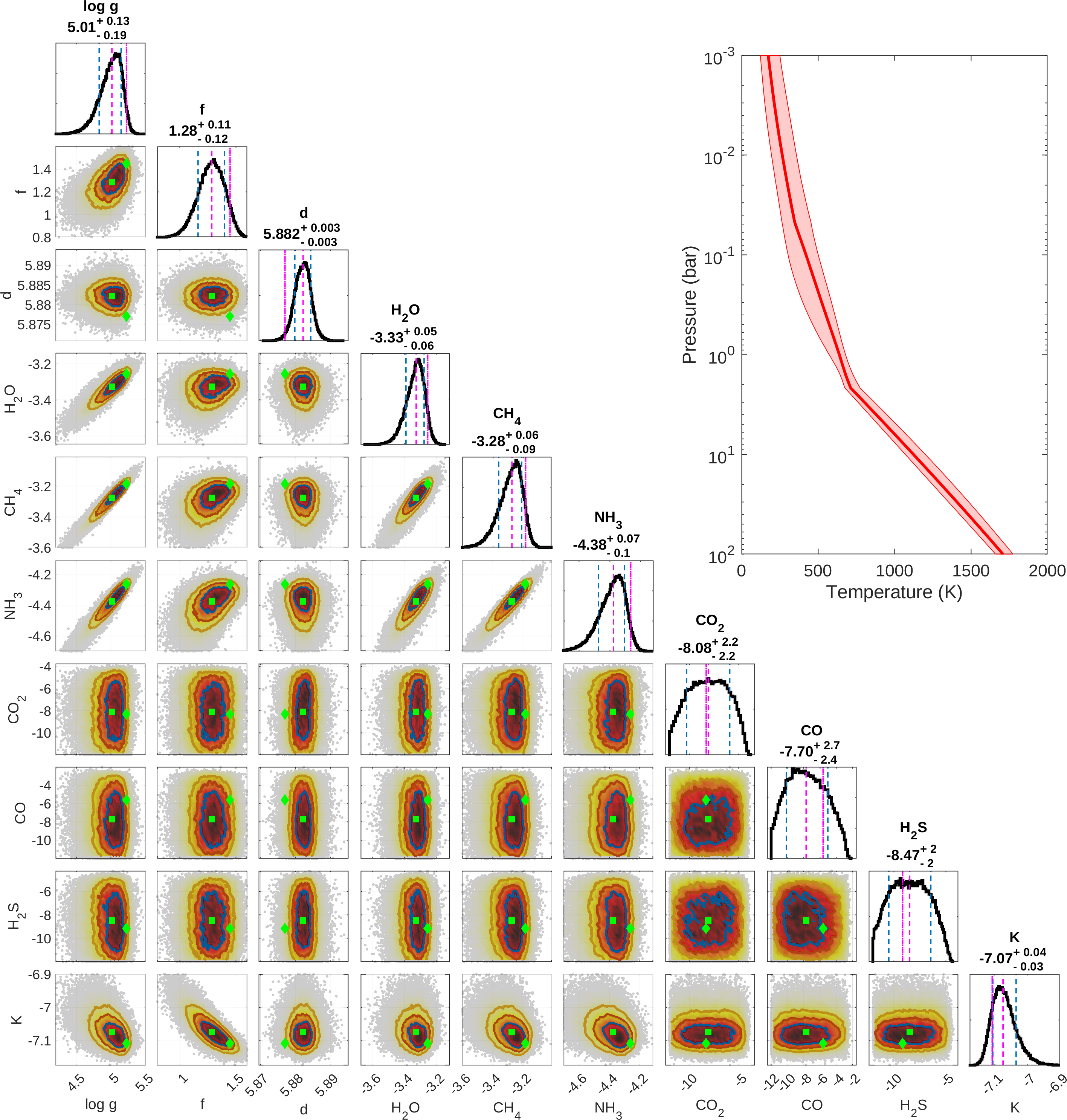}\vspace{-0.5cm}
	\end{center}
	\caption{Posterior distributions for the retrieval of GJ 570 D using a free chemistry approach (see Figure \ref{fig:helios_test3_post} for details on the posterior plots). The posteriors are depicted for the direct retrieval parameters. The plot in the upper right corner shows the retrieved temperature profile. The solid line corresponds to the median profile, while the shaded area indicates to the 1$\sigma$ confidence interval.}
	\label{fig:gj_retrieval_fc}
\end{figure*}

\begin{figure*}[t]
	\begin{center}
		\includegraphics[width=12cm]{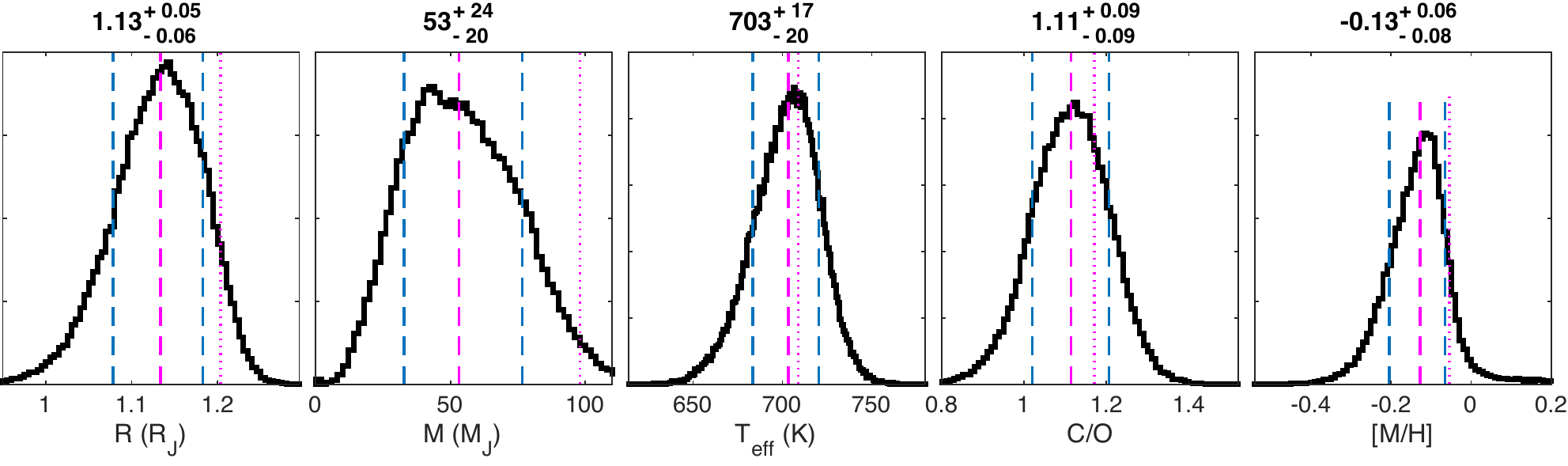}\vspace{-0.3cm}
	\end{center}
	\caption{Posterior distributions for the retrieval of GJ 570 D using a free chemistry approach. The posteriors are shown for the derived quantities: stellar radius $R$, stellar mass $M$, effective temperature $T_\mathrm{eff}$, and C/O ratio.}
	\label{fig:gj_retrieval_fc_deriv}
\end{figure*}

\begin{figure}[h]
	\begin{center}
		\includegraphics[width=\columnwidth]{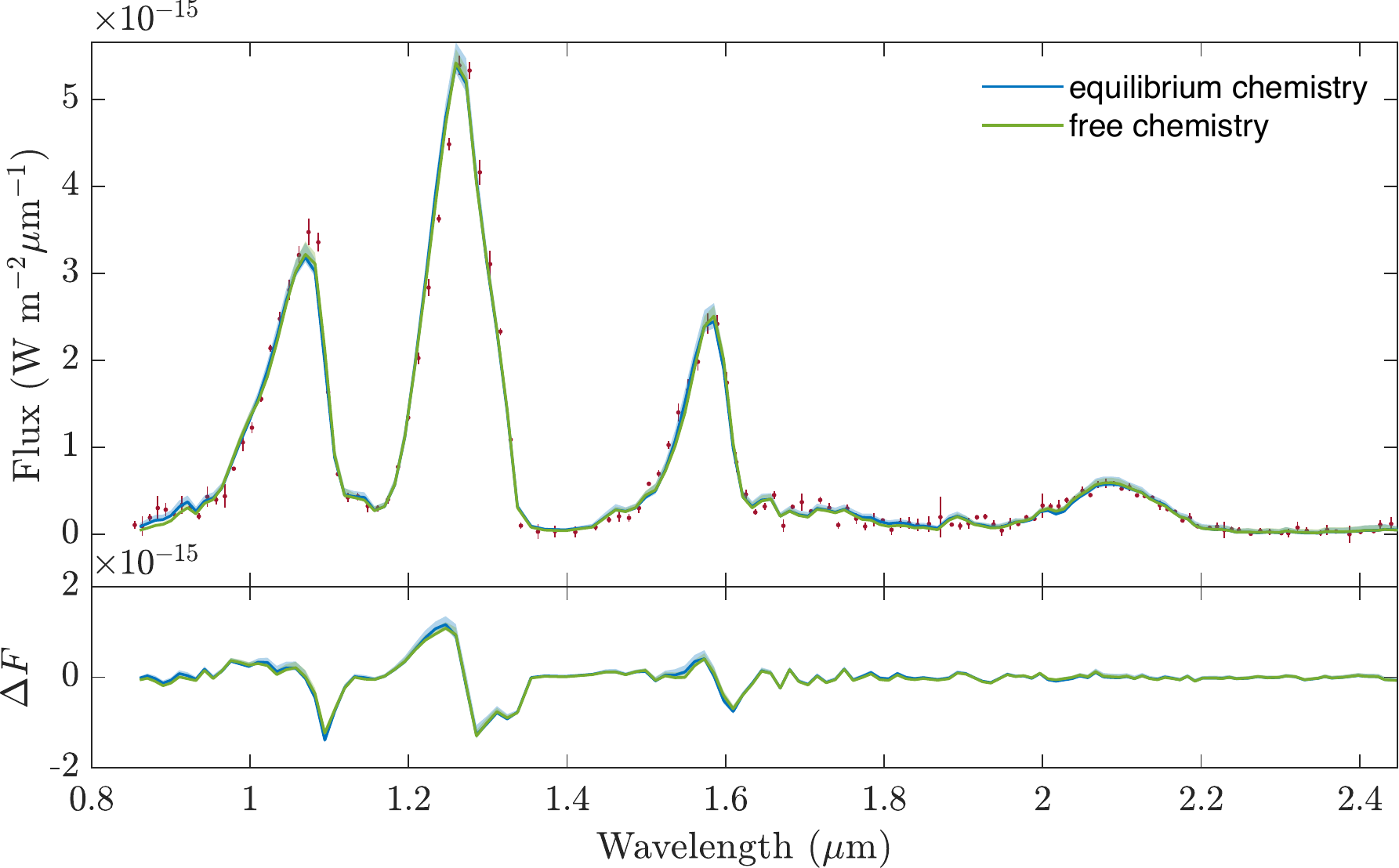}\vspace{-0.1cm}
	\end{center}
	\caption{Posterior spectra and residuals for the retrievals of GJ 570 D. The solid lines refer to the median of all posterior spectra for a retrieval with equilibrium chemistry (blue) and free chemistry (green). The shaded areas signify the 3$\sigma$ confidence intervals of the spectra. The measured spectrum of GJ 570 D is indicated by the red data points.}
	\label{fig:gj_retrieval_spectra}
\end{figure}

The results for \texttt{Helios-r2} with equilibrium chemistry are shown in Figure \ref{fig:gj_retrieval_eq}, while the ones with the free chemistry retrieval are depicted in Figures \ref{fig:gj_retrieval_fc} and \ref{fig:gj_retrieval_fc_deriv}. The corresponding spectra for both retrievals are presented in Figure \ref{fig:gj_retrieval_spectra}. 

Overall, most parameters seem to be quite well constrained in both cases. For the equilibrium chemistry forward model, we obtain a sub-solar metallicity of $-0.15^{+0.05}_{-0.04}$ and a super-solar C/O ratio of 0.83. The retrieved metallicity and C/O compares well with that of the host star reported by \citet{Line2015ApJ...807..183L}: $-$0.22--0.12 and 0.65--0.97, respectively. It should, however, be noted that our reported metallicity and C/O ratio corresponds to the pure gas phase and neglect the losses of elements due to condensation. The atmosphere's intrinsic, bulk element abundances, thus, might differ from our reported values.

Our retrieved value for the surface gravity of $4.61^{+0.08}_{-0.08}$ also matches the one reported by \citet{Line2015ApJ...807..183L} within their confidence intervals. When converting the retrieved $f$ parameter into a stellar radius via Eq. \eqref{eq:derived_distance}, we obtain a value of about 1 Jupiter radius. Combined with our $\log g$ value, we estimate a substellar mass of about $17^{+3.8}_{-3.0}$ $M_\mathrm{J}$, which is smaller than the one reported by \citet{Line2015ApJ...807..183L} ($31^{+24}_{-16}$ $M_\mathrm{J}$) but still contained within their confidence interval.

The free chemistry retrieval, on the other hand, yields slightly different results. With $5.01^{+0.13}_{-0.19}$ the retrieved $\log g$ is higher than for the equilibrium chemistry case, while the derived C/O ratio is about $1.11\pm0.09$, indicating an atmosphere that is enriched in carbon or, via condensation of oxygen-bearing species, depleted in O. The directly retrieved C/O ratio found by \citet{Line2015ApJ...807..183L} is 0.95--1.25 and, thus, well within our 1$\sigma$ confidence interval.

As suggested by the posterior distributions in Figure \ref{fig:gj_retrieval_fc}, we can constrain the abundances of water, methane, ammonia, and potassium. Upper bounds of roughly $10^{-5}$ are obtained for \ce{CO2} and \ce{H2S}, while an upper limit of $10^{-2}$ is found for \ce{CO}. When comparing the retrieved molecules' abundances with those from the free chemistry retrieval of \citet{Line2015ApJ...807..183L}, our model yields very similar median values.

Even though the value of the surface gravity is now higher than the one of the \citet{Line2015ApJ...807..183L} study, it is still consistent with those from other publications. As Table \ref{tab:retrieval_results} indicates, the spread in reported surface gravities for GJ 570 D extends from 4.5 to 5.23. Additionally, the $f$ factor, and thus the inferred radius, is larger than for the equilibrium chemistry case, which also results in a derived mass of about 53 Jupiter masses -- more than a factor of three higher than in the previous case.

Both retrievals also have similar posterior distributions for the effective temperature, both of which result in $T_\mathrm{eff}$ values of slightly larger than 700 K. This, again, is consistent with previous estimates by \citet{Line2015ApJ...807..183L} and \citet{Filippazzo2015ApJ...810..158F}, even though also higher temperatures have also been obtained by, e.g., \citet{delBurgo2009} and \citet{Testi2009} (see Table \ref{tab:retrieval_results}).

The temperature profiles obtained in both cases are quite similar, follow the form expected from the theory of brown dwarf atmospheres, and compare well to the one retrieved by \citet{Line2015ApJ...807..183L}. The smallest confidence intervals are found around pressures of 1 bar where most of the spectrum originates from. Larger intervals are obtained at lower and higher pressures. As mentioned in Sect. \ref{sec:temp_finite_elements}, our profiles are continuous and smooth by construction without any additional smoothing parameters required by other representations of the retrieved temperature profile.

Even though the results of the two different retrievals differ in terms of, e.g., surface gravity or metallicity, the spectra generated from the posterior distributions are surprisingly similar. As Figure \ref{fig:gj_retrieval_spectra} suggests, the median spectra only differ in details. For a larger part of the wavelength range, they are almost indistinguishable. Overall, the fit of the theoretical spectra to the actual observed one of GJ 570 D is quite good. The largest differences between the two are found at the 1.05 $\mu$m peak, where both of our retrievals have smaller values than the measured brown dwarf spectrum. This effect is also noticeable in the corresponding spectra shown in \citet{Line2015ApJ...807..183L}. It is possible that the description of the potassium far line wings from \citet{Allard2016A&A...589A..21A}, which have a large impact in this region, is still not satisfactorily representing the actual line shapes encountered in the atmospheres of brown dwarfs.

The resulting Bayesian evidence $\ln \mathcal Z_\mathrm{ec}$ of the equilibrium chemistry retrieval is 4775.73, while the free chemistry forward model yields a value of $\ln \mathcal Z_\mathrm{fc} = 4775.06$. The corresponding Bayes factor $B = Z_\mathrm{ec} / Z_\mathrm{fc}$ is 1.95. On the Jeffreys scale \citep{doi:10.1080/01621459.1995.10476572}, this indicates that there is no evidence to favor either one of the two different chemistry approaches. Both are equally likely to explain the data.

\subsection{Retrieval of Eps Indi Ba \& Bb}

In the following, we present our retrieval analysis of the two brown dwarfs in the $\epsilon$ Indi system, using the observational data from \citet{King2010AA...510A..99K}. In contrast to our retrieval of GJ 570 D from the previous subsection, we here have to impose an upper limit on the derived substellar masses of the brown dwarfs to obtain more realistic values for the retrieval parameters. Based on estimates of the hydrogen-burning main-sequence edge \citep{Burrows2001RvMP...73..719B}, we employ an upper mass limit of 80 $M_\mathrm{J}$. The same approach has also been used by, for example, \citet{Line2015ApJ...807..183L}.

\subsubsection{Equilibrium Chemistry Retrieval}

\begin{figure*}[ht]
	\begin{center}
		\includegraphics[width=\columnwidth]{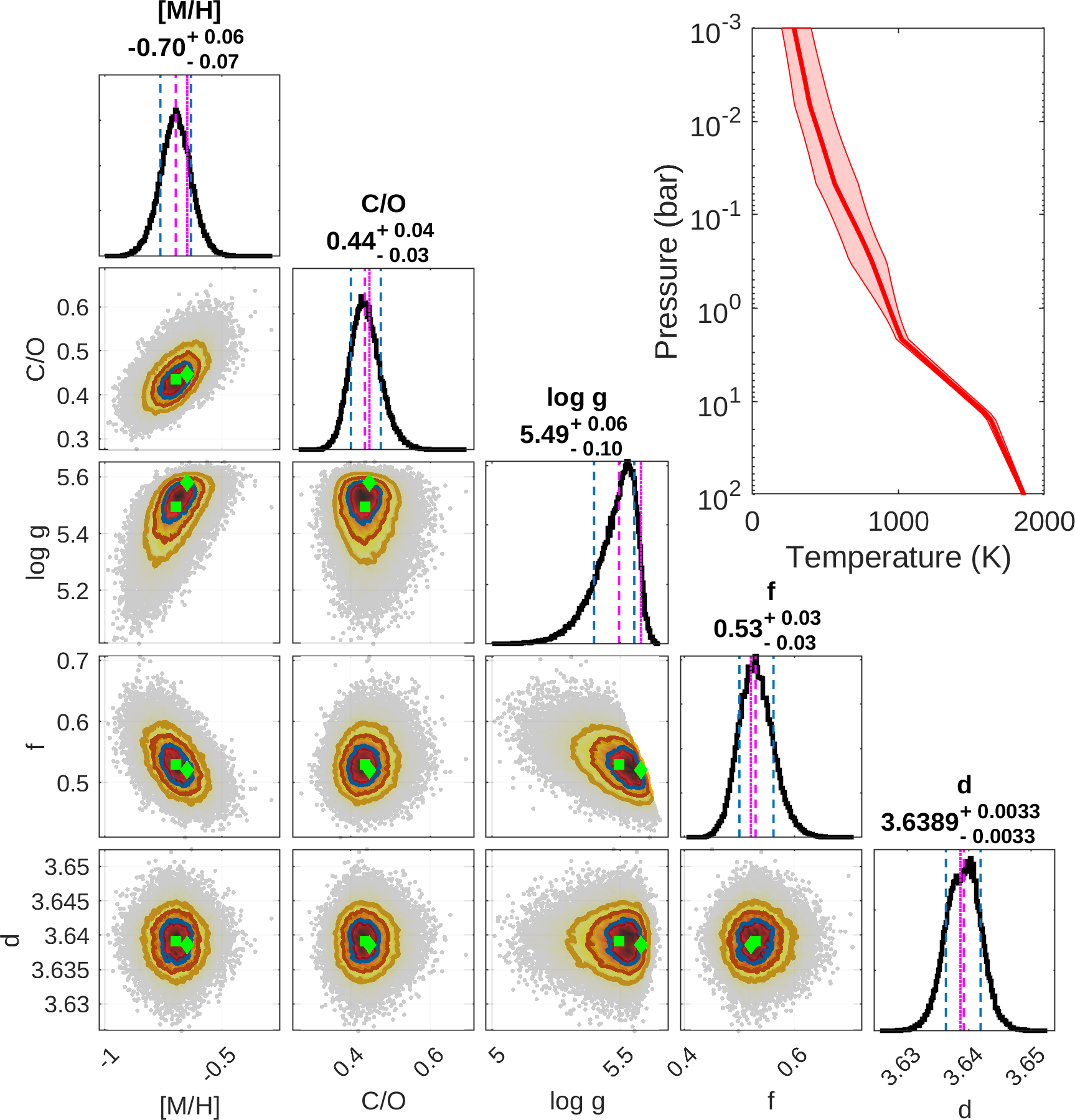} \includegraphics[width=\columnwidth]{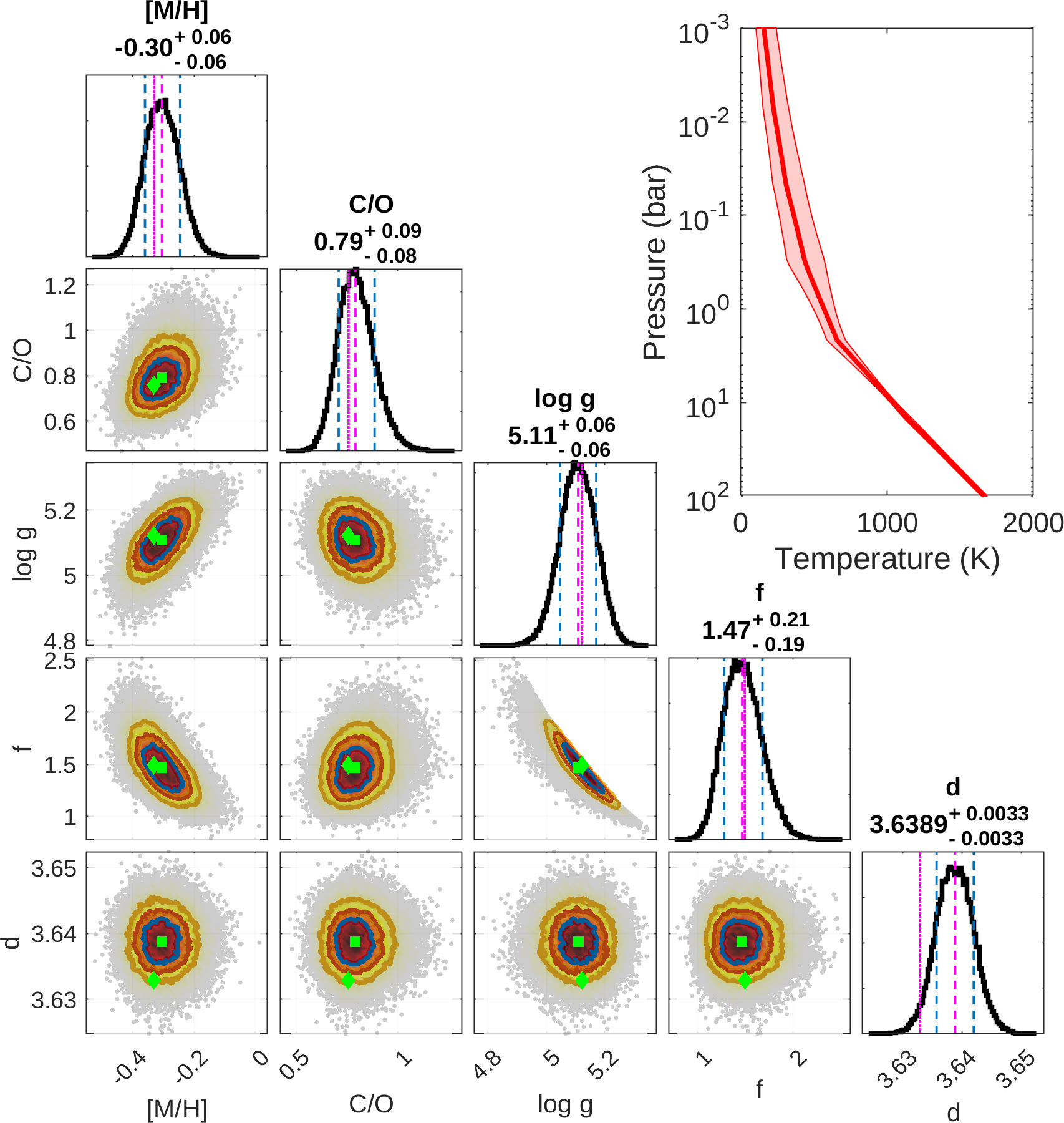}\vspace{0.2cm}
		\includegraphics[width=0.8\columnwidth]{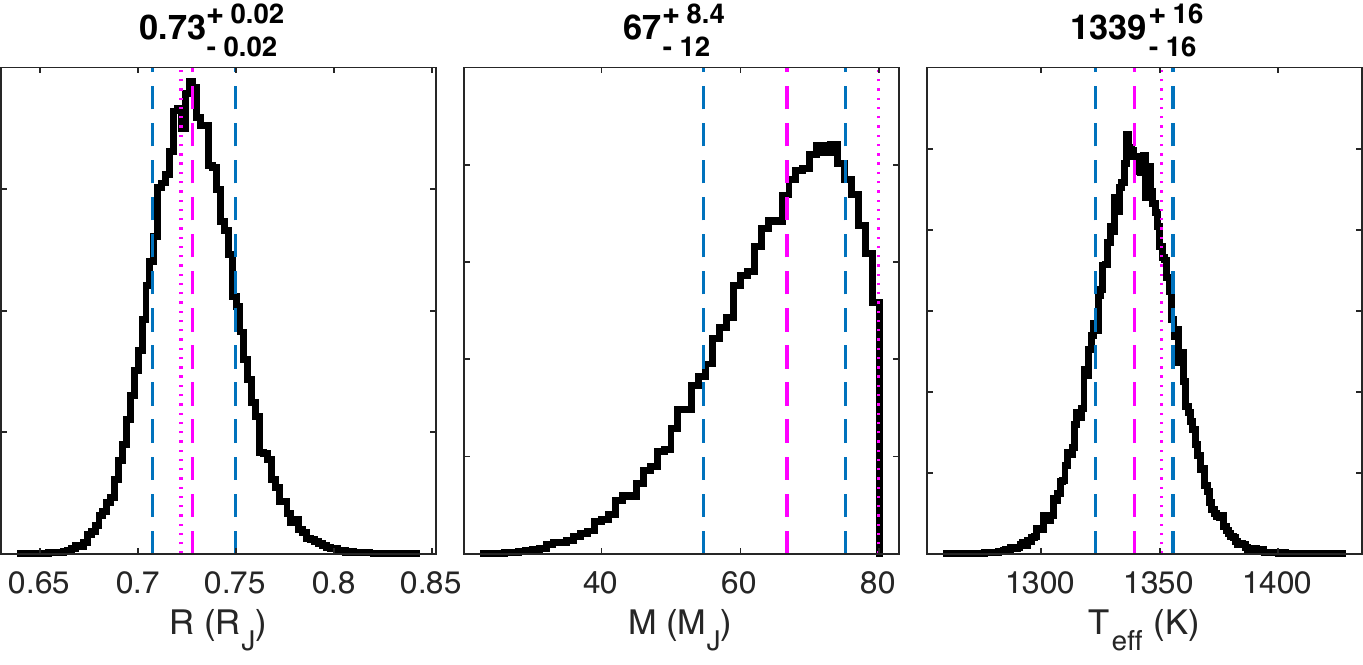} \hspace{1.5cm}  
    \includegraphics[width=0.8\columnwidth]{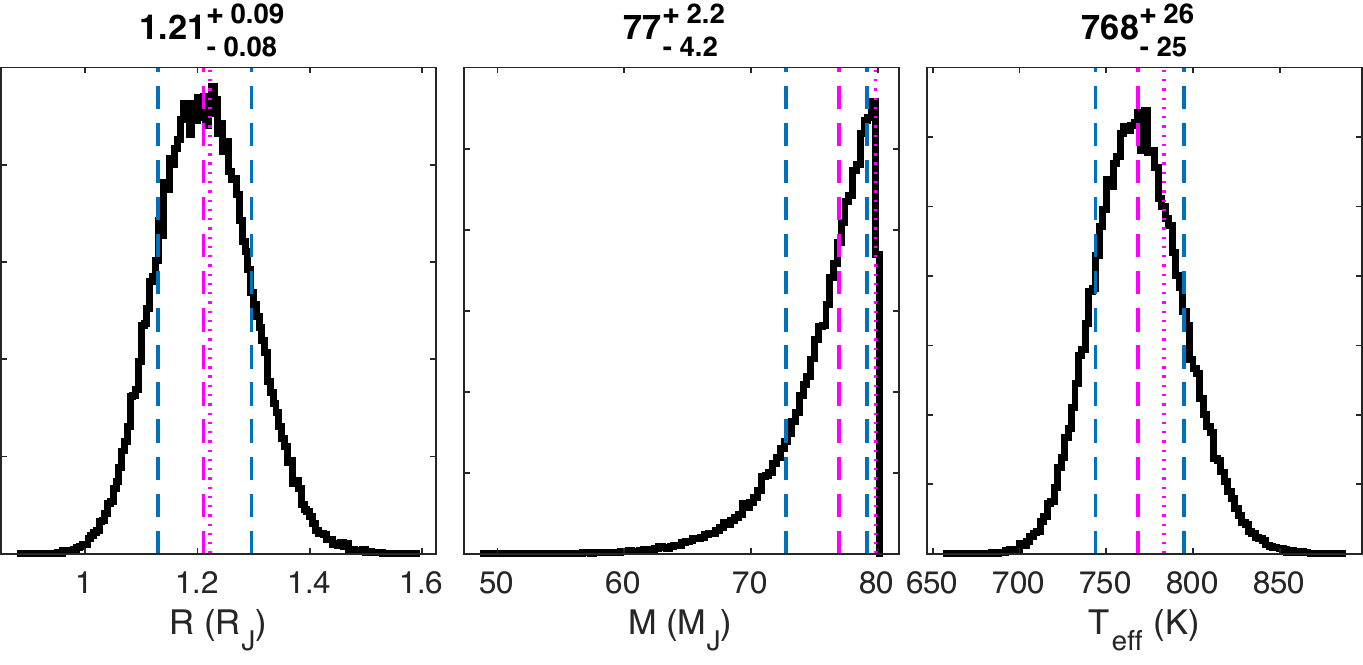} \vspace{-0.5cm}
	\end{center}
	\caption{Posterior distributions for the retrieval of $\epsilon$ Indi Ba \& Bb, employing an equilibrium chemistry approach (see Figure \ref{fig:helios_test3_post} for details on the posterior plots). The upper left panel shows the posteriors of the directly retrieved parameters for $\epsilon$ Indi Ba. The upper right panel the corresponding results for $\epsilon$ Indi Bb. The plots in the upper right corners show the retrieved temperature profiles. The solid lines correspond to the median profiles, while the shaded areas indicate the 1$\sigma$ confidence intervals. Posterior distributions for derived quantities are depicted in the lower panels.}
	\label{fig:epsind_retrieval_eq}
\end{figure*}

\begin{figure*}[ht]
	\begin{center}
		\includegraphics[width=0.9\textwidth]{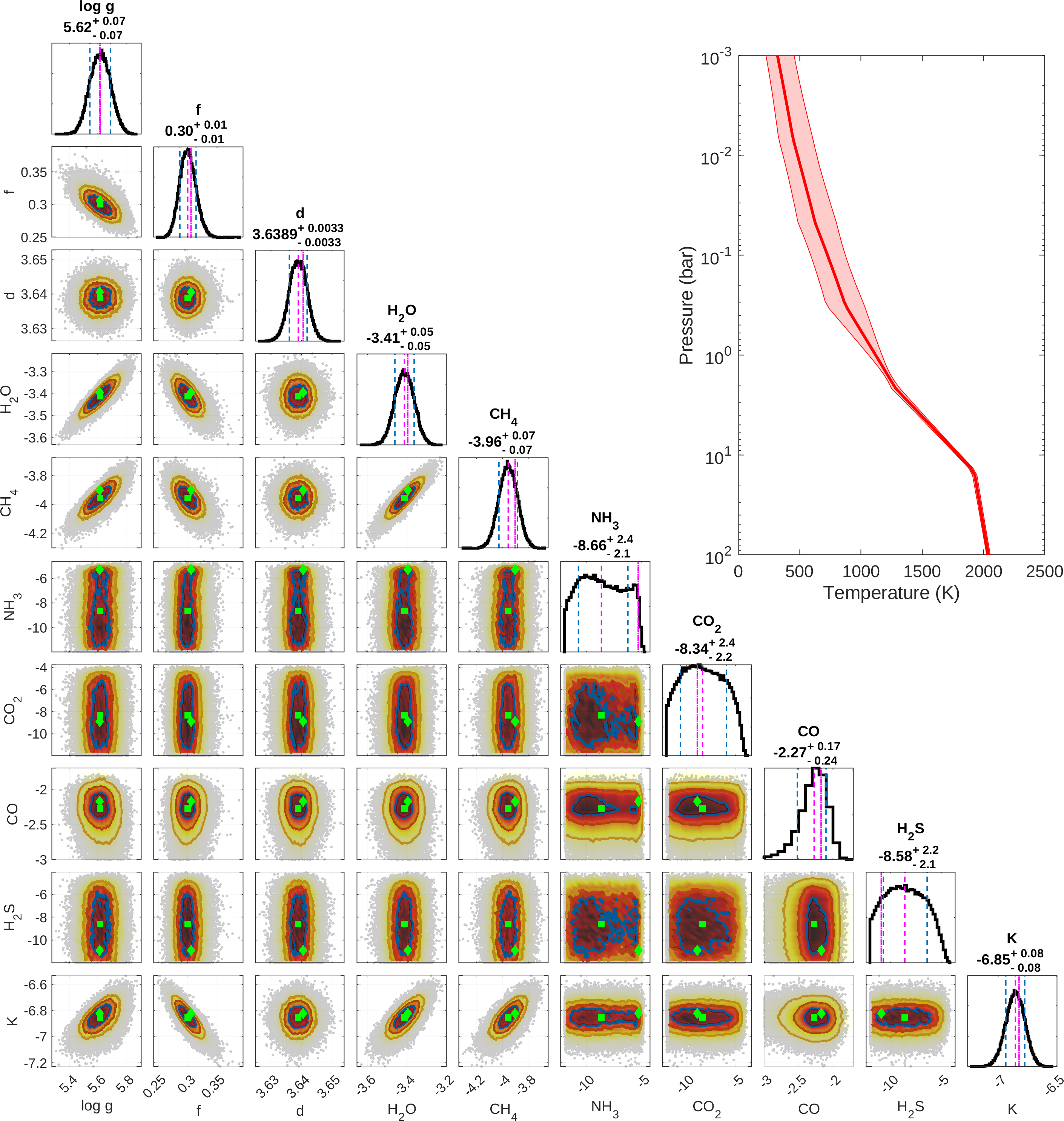}
	\end{center}
	\caption{Posterior distributions of directly retrieved parameters for the retrieval of $\epsilon$ Indi Ba, using a free chemistry approach (see Figure \ref{fig:helios_test3_post} for details on the posterior plots). The plot in the upper right corner shows the retrieved temperature profile. The solid line corresponds to the median profile, while the shaded area indicates the 1$\sigma$ confidence interval.}
	\label{fig:epsind_retrieval_fc}
\end{figure*}

The resulting posterior distributions for the equilibrium chemistry forward model are presented in Figure \ref{fig:epsind_retrieval_eq}, while those for free chemistry retrieval are shown in Figures \ref{fig:epsind_retrieval_fc} and \ref{fig:epsind_retrieval_fc_deriv}. The spectra and comparison to observations are given in Figure \ref{fig:epsind_retrieval_spectra}. A summary of the results and a comparison to the corresponding values from other studies is again presented in Table \ref{tab:retrieval_results}. We note that the potential impact of clouds is not accounted for in the retrieval calculations presented in this section. A corresponding retrieval for Eps Ind Ba with a gray cloud layer is shown in Appendix \ref{sect:app_cloud}.

In the case of equilibrium chemistry, we obtain quite different results for the overall metallicity of the two brown dwarfs. For $\epsilon$ Indi Ba, we retrieve a value of $-0.70\pm0.07$, while $\epsilon$ Indi Bb yields $-0.30\pm0.06$, respectively. With a C/O ratio of $0.44\pm0.04$ for $\epsilon$ Indi Ba and $0.79\pm0.09$ for $\epsilon$ Indi Bb, both are predicted to be enriched in oxygen compared to the solar value. However, just like the metallicity, the C/O ratios also differ by almost a factor of two. 

Another striking difference is the retrieved calibration parameter $f$. For the T1 dwarf, we obtain a value that is much smaller than unity, while for $\epsilon$ Indi Bb, the retrieved parameter is about  $1.47\pm0.21$. Since both brown dwarfs are part of the same system and, therefore, have the same distance to the observer, the distinctively different $f$ factor cannot originate from an erroneous distance estimate. The very different predicted values for $f$ also result in inferred radii that differ by quite a wide margin. The radius of $\epsilon$ Indi Ba is estimated to be 0.73 Jupiter radii, which is smaller than one would expect from a brown dwarf, while our results for $\epsilon$ Indi Bb infer a radius of 1.21 $R_\mathrm{J}$. As mentioned earlier, the very low, retrieved radius for the early T dwarf could also be the result of a heterogeneous atmosphere that has a smaller effective emitting area. Alternatively, $f$ might also again compensate for missing or oversimplified model physics.

Our retrieved surface gravities for both brown dwarfs are also quite high. In particular, the value of $5.49^{+0.06}_{-0.10}$ for $\epsilon$ Indi Ba is higher than those reported by most previous studies (see Table \ref{tab:retrieval_results}). For $\epsilon$ Indi Bb, we obtain a value of $5.11\pm0.06$ which, on the other hand, agrees very well with previous estimates. It should, however, be noted that these values are influenced by the restriction of our retrievals to a total derived mass of 80 Jupiter masses, which is thought to be the upper mass limit for brown dwarfs before the star becomes heavy enough to ignite the hydrogen burning in its core. As can be clearly noted in the correlation plots of Figure \ref{fig:epsind_retrieval_eq}, the posteriors for $f$ and $\log g$ are cut at higher values. A fully free retrieval would probably have resulted in even higher values of the surface gravity. Dynamical masses for the two companions have been reported by \citet{dieterich18}. With 75 $M_\mathrm{J}$ ($\epsilon$ Indi Ba) and 70.1 $M_\mathrm{J}$ ($\epsilon$ Indi Bb), respectively, these masses also quite high. At least for $\epsilon$ Indi Ba, the measured mass is contained within the retrieved 1$\sigma$ confidence interval.

The same also applies to the inferred masses (see bottom panel of Figure \ref{fig:epsind_retrieval_eq}). The posterior for $\epsilon$ Indi Bb is clearly cut at the upper mass limit, while the one for $\epsilon$ Indi Ba is also skewed toward higher values of $M$.

Our derived equilibrium temperature of about 1339 K in case of the T1.5 dwarf $\epsilon$ Indi Ba falls within the predicted range of 1250 K -- 1400 K published in earlier studies. For $\epsilon$ Indi Bb, a T6 brown dwarf, we obtain a value of 768 K that is cooler than the lower bound (875 K) estimated by \citet{Kasper2009ApJ...695..788K}.

\begin{figure*}[ht]
	\figurenum{\ref{fig:epsind_retrieval_fc} cont}
	\begin{center}
		\includegraphics[width=0.9\textwidth]{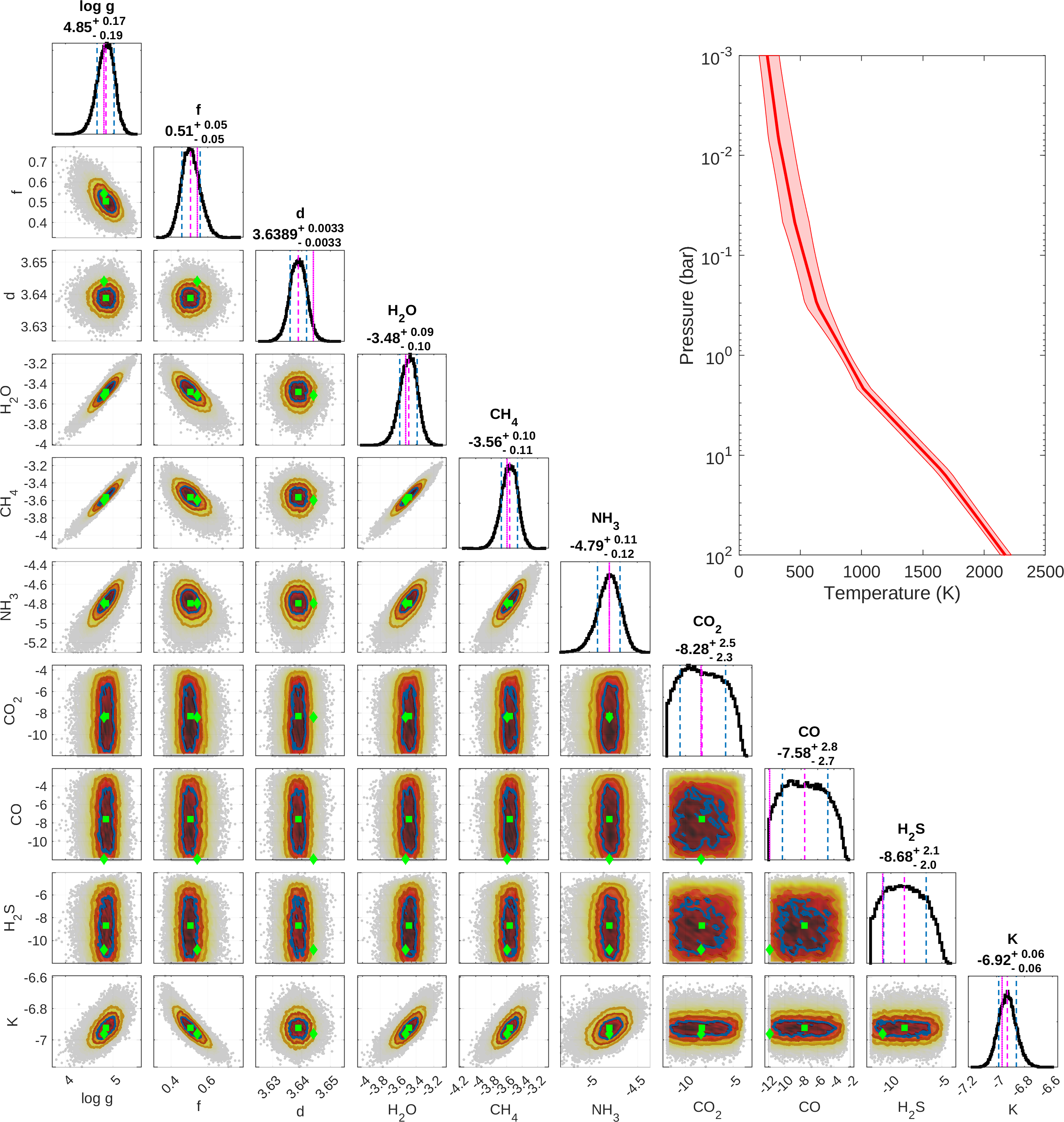}
	\end{center}
	\caption{Same as above but for $\epsilon$ Indi Bb.}
\end{figure*}

One striking difference between the two brown dwarfs is the retrieved temperature profiles. The T1 dwarf $\epsilon$ Indi Ba shows a peculiar, shallow lapse rate in the lower atmosphere, which is absent in both later T dwarfs, GJ 570 D and $\epsilon$ Indi Bb, of our study. Such shallow profiles are normally not expected from the standard brown dwarf atmosphere models. They usually predict much steeper lapse rates in the lower atmosphere that are either given by the dry/moist adiabates or a temperature profile in radiative equilibrium \citep{Marley2015ARA&A..53..279M}. This behavior might be caused by the lack of an isothermal cloud layer in the lower atmosphere in the current retrieval model.

\subsubsection{Free Chemistry Retrieval}

\begin{figure*}[t]
	\begin{center}
		\includegraphics[width=12cm]{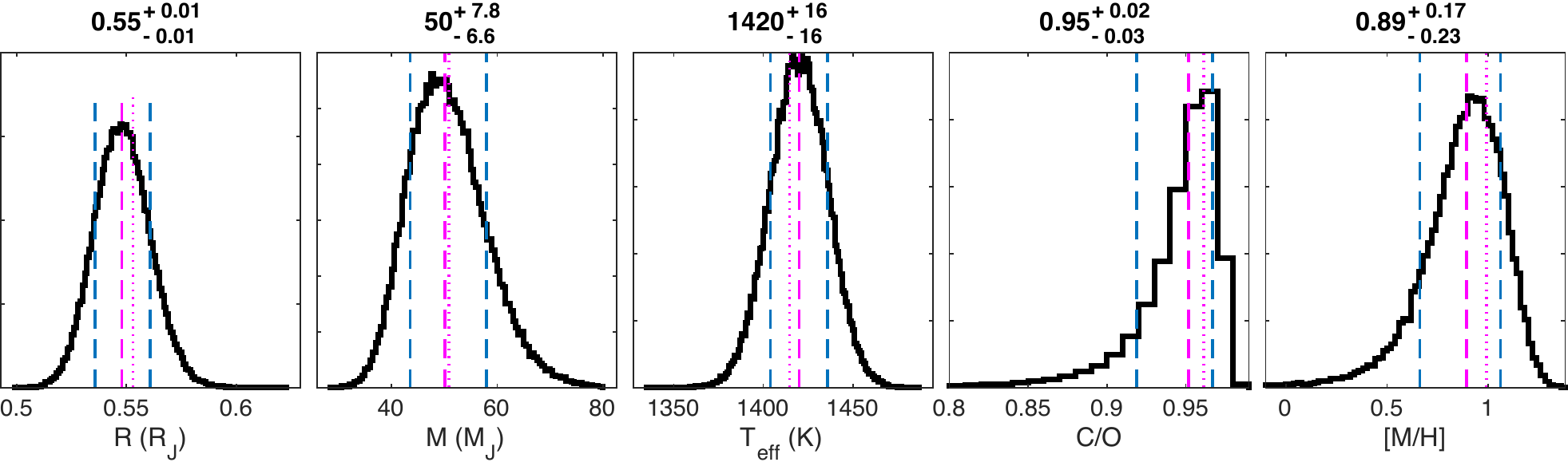}
		\includegraphics[width=12cm]{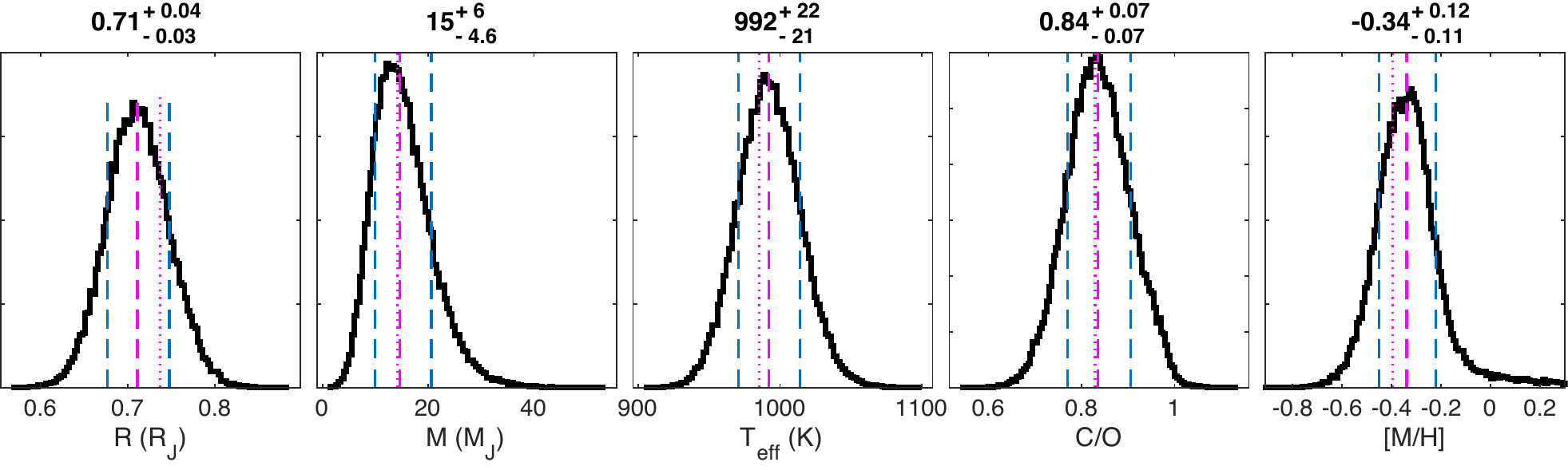} \vspace{-0.3cm}
	\end{center}
	\caption{Posterior distributions for the retrievals of $\epsilon$ Indi Ba (upper panel) and $\epsilon$ Indi Bb (lower panel) using a free chemistry approach. The posteriors are shown for the derived quantities: stellar radius $R$, stellar mass $M$, effective temperature $T_\mathrm{eff}$, and C/O ratio.}
	\label{fig:epsind_retrieval_fc_deriv}
\end{figure*}

The results for the free chemistry retrievals are shown in Figures \ref{fig:epsind_retrieval_fc} and \ref{fig:epsind_retrieval_fc_deriv}.

Just like the results for the equilibrium chemistry case, the retrieved values for the calibration parameter $f$ are well below of its expected value of around unity. For $\epsilon$ Indi Ba, we now obtain the very low value of $0.30\pm0.01$, which results in an inferred radius of just 0.55 Jupiter radii. Such a result might be unphysical for a homogeneously emitting atmosphere. On the other hand, this may also reflect a heterogeneous atmosphere. Heterogeneities would result in a reduction of the effective emitting area and thus yield a smaller than expected radius.
The corresponding retrieved value for $\epsilon$ Indi Bb is now smaller than unity and with $0.51\pm0.05$ (inferred radius 0.71 Jupiter radii). Unlike the equilibrium chemistry case, the posteriors for $\log g$ and $M$ are not affected by the upper mass limit of 80 Jupiter masses. The surface gravity posterior we obtain for $\epsilon$ Indi Ba is a bit higher than its equilibrium chemistry value.
However, for $\epsilon$ Indi Bb, we now obtain a much lower value of $4.85^{+0.17}_{-0.19}$, which is still roughly consistent with the lower end of previously published values (see Table \ref{tab:retrieval_results}). Compared to the equilibrium chemistry retrieval, the inferred mass of $\epsilon$ Indi Bb is now much lower ($\approx 15 M_\mathrm{J}$) but also deviates strongly from the dynamical mass of $\approx 70 M_\mathrm{J}$ predicted by \citet{dieterich18}.

In the case of $\epsilon$ Indi Ba, we obtain estimates on the abundances of \ce{H2O}, \ce{CH4}, \ce{CO}, and \ce{K}. The ones for \ce{NH3}, \ce{H2S}, and \ce{CO2} are unconstrained.
Carbon monoxide is predicted to be more abundant than \ce{CH4} which is possible for an object close to the L-T transition. On the other hand, one would expect \ce{NH3} to be largely absent, which is confirmed by our results. Compared to the retrieved sub-solar C/O ratio for the equilibrium chemistry case, we now obtain a derived posterior mean value of $0.95\pm0.03$ which suggests a super-solar composition in terms of C/O. As already discussed for GJ 570 D, this C/O ratio is affected by condensation of oxygen-bearing condensates. The bulk C/O ratio of the atmosphere is most likely smaller than predicted by the retrieved abundances because of oxygen atoms locked in condensates.

Since $\epsilon$ Indi Bb is a cooler T6.5 dwarf, the atmosphere is more enriched in methane than in \ce{CO}. In this case, we obtain constraints on \ce{H2O}, \ce{CH4}, \ce{NH3}, and \ce{K}, while \ce{CO}, \ce{CO2}, and \ce{H2S} are unconstrained. In fact, the results are roughly similar to the free chemistry ones from GJ 570 D (see Figure \ref{fig:gj_retrieval_fc}). One distinct difference is that in the latter case, \ce{CH4} is predicted to be more abundant than water, while for the former brown dwarf, \ce{H2O} is the more abundant molecule. Consequently, the derived C/O ratio for $\epsilon$ Indi Bb is still smaller than unity (but super-solar), while for GJ 570 D, we obtain a carbon-rich atmosphere.

The derived equilibrium temperature for $\epsilon$ Indi~Ba of 1420 K is still within the range of previously published values (see Table \ref{tab:retrieval_results}). The value of 992~K we obtain for the free chemistry case of $\epsilon$ Indi~Bb, is close to the upper bound of 940~K published by \citet{King2010AA...510A..99K}.

Just like in the case of the equilibrium chemistry model, we again obtain the very shallow temperature profile in the lower atmosphere of $\epsilon$ Indi Ba. In fact, the temperature profile seems to show an even stronger gradient than found in the previous case. For the late-T dwarf $\epsilon$ Indi Bb, we also obtain a very similar profile as before.

The logarithmic Bayes factors $\ln B = \ln Z_\mathrm{fc} - \ln Z_\mathrm{ec}$, based on the Bayesian evidences for the free ($\mathcal Z_\mathrm{fc}$) and the equilibrium chemistry ($\mathcal Z_\mathrm{ec}$) are 34.67 ($\epsilon$ Indi~Ba) and 59.06 ($\epsilon$ Indi~Bb), respectively. On the Jeffreys scale, the free chemistry model is therefore favored decisively over the equilibrium chemistry one. 

\subsubsection{Comparison of Spectra}

\begin{figure}[ht]
	\begin{center}
		\includegraphics[width=\columnwidth]{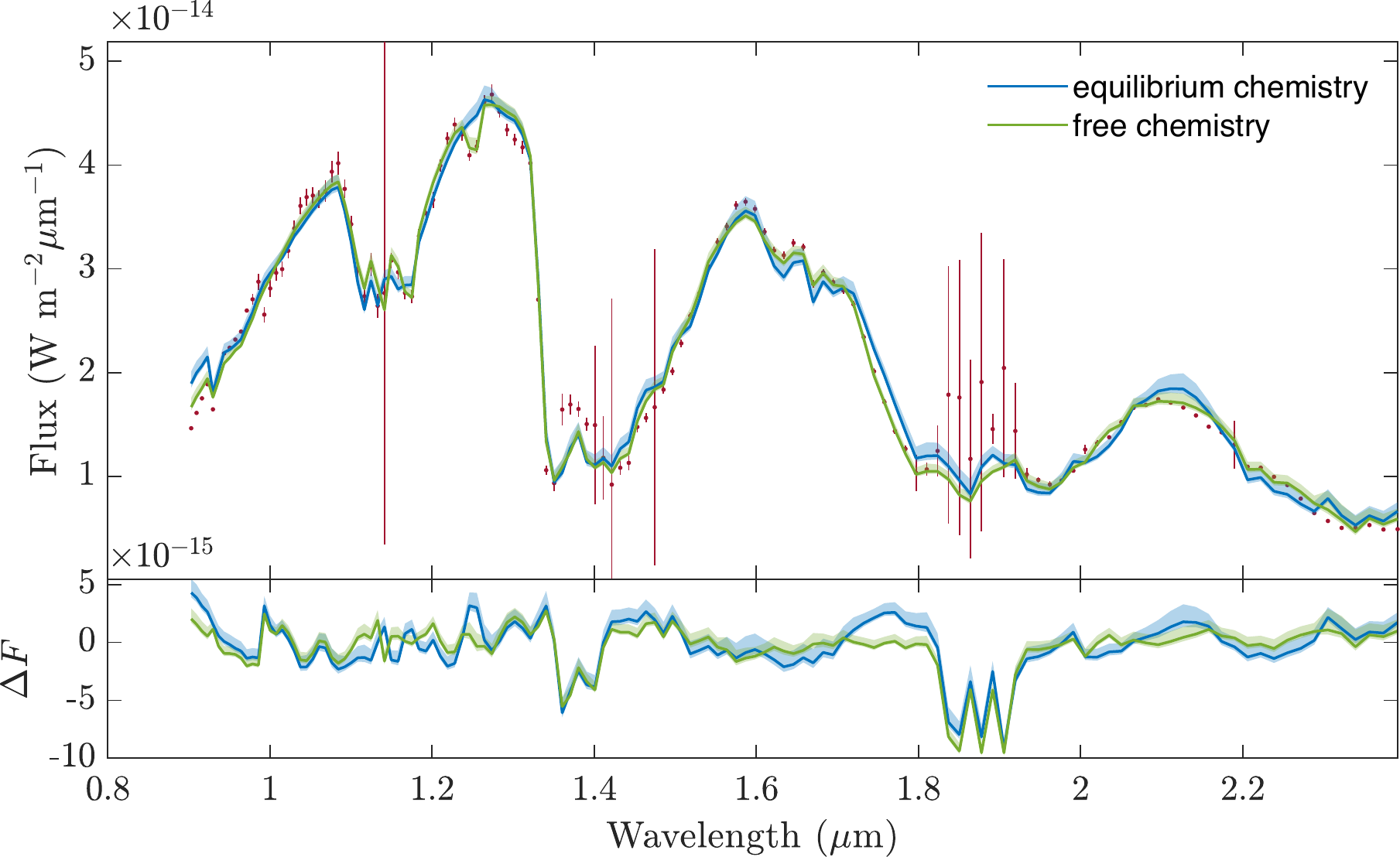}
		\includegraphics[width=\columnwidth]{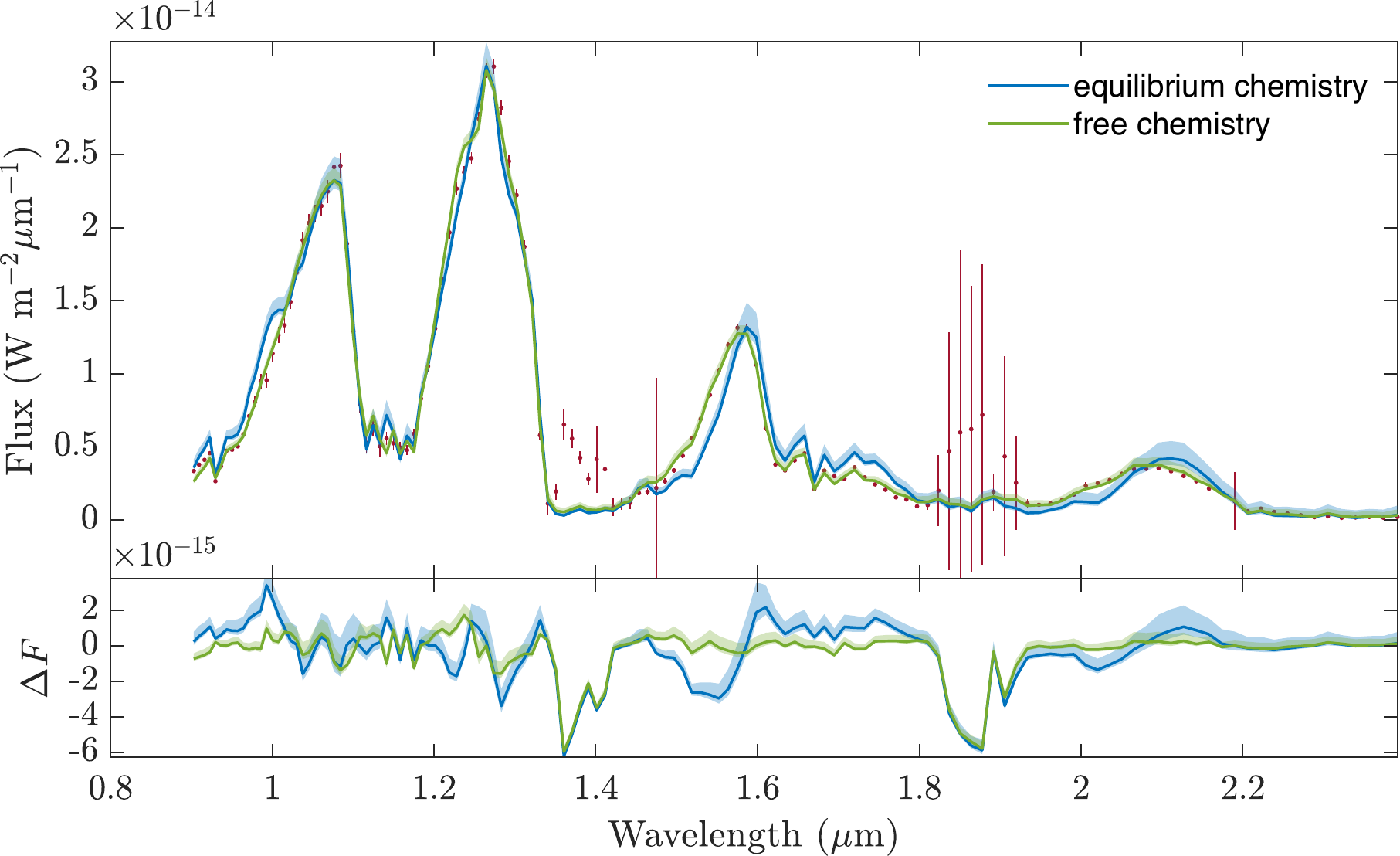}
	\end{center}
	\caption{Posterior spectra and residuals for the retrievals of $\epsilon$ Indi Ba (upper panel) and $\epsilon$ Indi Bb (lower panel). The solid lines refer to the median of all posterior spectra for a retrieval with equilibrium chemistry (blue) and free chemistry (green). Shaded areas signify the 3$\sigma$ confidence intervals of the spectra. The measured spectra of $\epsilon$ Indi Ba and $\epsilon$ Indi Bb are indicated by the red data points.}
	\label{fig:epsind_retrieval_spectra}
\end{figure}

Figure \ref{fig:epsind_retrieval_spectra} depicts the posterior spectra for both chemistry retrievals of the brown dwarfs in the $\epsilon$ Indi system. The measured spectra are also shown for a direct comparison.

The resulting spectra clearly suggest that the chemical equilibrium model is not able to fully reproduce the measured spectra. Larger deviations can be noticed at wavelengths around 2.1 $\mu$m, where the peak in the measured spectrum is both overpredicted and shifted to larger wavelengths.
These discrepancies could be caused by a pressure-dependent H$_2$ absorption effect based on the different temperature profiles. The emission for the equilibrium chemistry might originate from slightly higher pressures and, thus, is impacted by a higher \ce{H2} continuum absorption. Other possibilities are a mix of overlapping \ce{H2O} and \ce{CH4} absorption or, considering the fact that this discrepancy does not occur for GJ 570 D, instrument systematics.
In the case of $\epsilon$ Indi Bb (lower panel), further deviations can be seen at the peaks near 1.4 and 1 $\mu$m. For $\epsilon$ Indi Ba (upper panel), one can see a striking difference between the two different chemistry approaches in the feature at 1.25 $\mu$m.
Overall, the free chemistry approach yields an excellent fit to the data. Larger differences compared to the data can again be seen in the peak near 1 $\mu$m for both brown dwarfs. As explained already for GJ 570 D, these differences most likely originate from the description of the line-wing profiles of potassium. The only regions that cannot be explained by any model are the elevated flux values at about 1.6 and 1.9 $\mu$m. \citet{King2010AA...510A..99K} attributed these features to an unknown absorber. However, given the large error bars for most of the data points in these regions, these features could also be caused by, e.g., measurement or post-processing errors.

\section{Discussion}
\label{sect:discussion}

As discussed in the previous section, we obtain overall quite different results when assuming equilibrium chemistry or by using the free chemistry approach. While in the case of GJ 570 D, both approaches yield results that are similar, for the two brown dwarfs in the $\epsilon$ Indi system, the results seem to be remarkably different. Additionally, based on the evaluation of the Bayes factor between the two different approaches, the equilibrium chemistry and the free chemistry approach are equally likely to explain the data in case of GJ 570 D. For the Epsilon Indi brown dwarfs, on the other hand, equilibrium chemistry is decisively disfavored compared to the free chemistry mode. This might suggest that despite their comparable stellar classification as late-T dwarfs, the atmospheres of GJ 570 D and $\epsilon$ Indi Ba are chemically quite distinct.

Most atmosphere model grids that were used to analyze brown dwarf data so far (e.g. \citet{allard01}, M.~S. Marley et al., 2020 in preparation) have exclusively used equilibrium chemistry, with or without a treatment of condensation. Retrievals (e.g., \cite{Line2015ApJ...807..183L}), on the other hand, usually employ a free chemistry approach. This larger number of additional free parameters gives the retrieval usually more freedom to fit the spectrum properly. In contrast to a free chemistry that retrieves individual mixing ratios for each constituent, equilibrium chemistry has only two free parameters: the metallicity ([M/H]) and C/O ratio. Using this approximation can strongly limit the flexibility of a retrieval model.
While a free chemistry retrieval seems to provide mostly a better fit to the data, it has to assume that the mixing ratios of the chemical species are isoprofiles. This, of course, is not expected to be the case in any real atmosphere.

In principle, it is also possible to give the equilibrium chemistry more flexibility by softening the assumption that the ratios of the element abundances (except for C and O) are solar and allow the actual element abundances to change freely. We will explore this issue in more detail within a future study.\\

Reporting derived radii and masses seems to be difficult. As pointed out in Sect. \ref{sec:radius_distance_relation}, the inferred radii are all based on the calibration factor $f$, which, however, assumes that $f$ only includes information on the radius. In practice, though, $f$ encompasses different error sources, such as errors in the photometric calibration, errors in the distance measurement that have not been accounted for in the prior for $d$, a reduced emitting area due to a heterogeneous atmosphere, or even model inadequacies. The inferred radii and masses, thus, should be taken with great caution.

One possible way to overcome this issue is the direct measurement of brown dwarf radii. This, for example, is possible via long-baseline infrared interferometry in the future \citep{Burgasser2019BAAS...51c.214B} or by measuring the transit depths of eclipsing brown dwarfs \citep{Dupuy2009ApJ...704.1519D}.

As we have shown in the Sect. \ref{sec:helios_test}, this parameter can also be used by the retrieval model to account for differences in the alkali line-wing descriptions. In cases where \texttt{Helios-r2} was used to retrieve a spectrum from the \texttt{Sonora} grid, including $f$ as a free parameter provides a better fit to the spectrum near 1 $\mu$m but leads to surface gravities that are too high and metallicity as well as C/O ratios that differ from the actual ones. Obtaining derived radii (and thus masses) from this retrieved parameter alone might therefore lead to misleading results.

In a companion paper \citep{Oreshenko2020AJ....159....6O}, we explore the same three brown dwarfs by using different model grids obtained from self-consistent brown dwarf atmosphere models in combination with a random forest machine-learning approach.
For the $\epsilon$ Indi system, we also obtain very low $f$ factors, indicating that perhaps the photometric calibration performed by \citet{King2010AA...510A..99K} contains systematic errors. We conclude that maybe an independent calibration of the data or additional measurements would help to address this issue in the future.\\

Our retrieved temperature profile for the early T dwarf $\epsilon$ Indi Ba shows a peculiar, shallow lapse rate in the lower atmosphere that we do obtain for the later T dwarfs GJ 570 D or $\epsilon$ Indi Bb. Such a result is usually not found in temperature profiles obtained by the usual self-consistent brown dwarf models \citep[e.g.][]{Burrows1993ApJ...406..158B, allard01, saumon08}.
This retrieved profile could be the result of missing model physics, in particular, cloud layers that might be present in the photosphere of the T1-type brown dwarf and absent for the two late-T dwarfs. This type of behavior for retrieved temperature profiles has been discussed by \citet{Burningham2017MNRAS.470.1177B}, for example. Essentially, a cloud layer would block the light from the deeper, hotter regions. As a result, a cloud-free retrieval of a cloudy atmosphere would try to mimic this behavior by reducing the lower atmospheric temperatures and producing a more isothermal profile at higher pressures. In the Appendix \ref{sect:app_cloud} we perform a retrieval of $\epsilon$ Indi Ba with an additional gray cloud layer. The resulting cloud parameters are essentially unconstrained and the other posterior distributions are equal to those of the cloud-free case. The results, thus, suggest that the absence of a gray cloud layer is unable to explain the shallow lapse rate in the lower atmosphere.

Another possible scenario that can create such temperature profiles is based on the idea of a thermo-chemical instability. In a series of publications, \citet{Tremblin2015ApJ...804L..17T} and \citet{Tremblin2016ApJ...817L..19T} argued that the L-T transition is caused by a fingering convective instability rather than due to cloud layers. They propose a super-adiabatic lapse rate in the lower atmospheres of brown dwarfs at the L-T transition, comparable to the one that we retrieve for the T1 dwarf $\epsilon$ Indi Ba.

Finally, it is also possible that such a profile is the result of applying a simple one-dimensional model to an inherent three-dimensional object. The actual, measured spectrum of the brown dwarf is a convolution of emitted light from different parts of the visible hemisphere. These parts do not necessarily share the same temperature profile, chemistry, or cloud coverage. Indeed, rotational modulations are commonly seen in brown dwarfs \citep[][]{Apai2014ApJ...782...77B,Metchev2015}  and are thought to be caused by spatial variations of cloud thickness and temperature \citep[e.g.,][]{Radigan2012,Apai2013,Buenzli2015}, probably driven by atmospheric circulation \citep{Apai2017Sci...357..683A,Showman2018,TanShowman2019}. These studies found different pressure-dependent phase offsets in multiwavelength spectrophotometry between L/T transition \citep{Apai2013,Yang2016} and late-T dwarfs \citep{Buenzli2012,Yang2016}, possibly suggesting a different atmospheric structure. These findings show that brown dwarfs are not simple, spatially homogeneous objects and that applying a single, one-dimensional model to retrieve physical quantities might yield unexpected results.

In general, applying a model like \texttt{Helios-r2} to retrieve spectra of brown dwarfs or exoplanets should always involve a hierarchy of models with different assumptions or modeling approaches (e.g. equilibrium chemistry vs. free chemistry retrieval) and use the resulting Bayesian evidences to perform a model selection (see \citet{Lavie2017AJ....154...91L}). The complexity of the temperature profile parameterization used in the retrieval should be chosen to match the available data quality in terms of wavelength coverage or spectral resolution. For example, choosing a high-order, piecewise polynomial with many elements for an exoplanet spectrum taken with the Wide Field Camera 3 on the \textit{Hubble Space Telescope} that typically has about 10 to 12 data points will most certainly lead to overfitting. To ensure, that the temperature profile is retrieved correctly, convergence tests should ideally be made. Retrieving a more complex temperature profile than the ones studied here (e.g. temperature inversions) usually require a higher number of elements, as we demonstrate in \citet{Bourrier2019arXiv190903010B}.

\section{Summary}
\label{sect:summary}

In this work, we present our new retrieval model \texttt{Helios-r2}. The code is open source and available on our Exoclimes Simulation Platform\textsuperscript{\ref{fnote:EEG}}.
Compared to the previous version used in \citet{Lavie2017AJ....154...91L} and \citet{Oreshenko2017ApJ...847L...3O}, \texttt{Helios-r2} has been completely rewritten. It includes the option to directly use a complex equilibrium chemistry model during the parameter space exploration as well as performing free chemistry retrievals, if desired. Furthermore, we add a novel representation of the temperature profile based on piecewise polynomials comparable to a finite element approach. This allows for free-temperature retrievals that yield smooth, continuous temperature profiles without requiring additional smoothing parameters. Additionally, \texttt{Helios-r2} can also retrieve T-p profiles based on Milne's solution, as used by the previous version \citep{Lavie2017AJ....154...91L}. 
The possibility of using optional instrument profiles to simulate observed spectra is now available as well. The retrieval model uses a Bayesian approach by employing a nested-sampling parameter space explorer. It provides the Bayesian evidence as well as the posterior distributions of the retrieval parameters. 

As a first test, we apply \texttt{Helios-r2} to three brown dwarf atmospheres: the late-T dwarfs GJ 570 D and $\epsilon$ Indi Bb, as well as the T1 dwarf $\epsilon$ Indi Ba.
We retrieve the chemical composition, temperature profiles, and  derive surface gravities, radii, and effective temperatures. For GJ 570 D, our results agree well with previous estimates. Our retrieved $\log g$ values for $\epsilon$ Indi Ba \& Bb are broadly consistent with those inferred from dynamical masses. 
The radii of both $\epsilon$ Indi brown dwarfs, derived from the retrieved calibration factor $f$, have smaller values than expected for these types of objects. Because $f$ does not only describe the radius but also includes other potential sources of error as well as inadequate model physics, the inferred radii and masses, thus, should be taken with great caution.
For the two brown dwarfs of the $\epsilon$ Indi B system, the solutions resulting from a free chemistry retrieval are favoured over the equilibrium chemistry approach and provide a better fit to the measured spectra. For GJ 570 D, on the other hand, equilibrium chemistry and a free chemistry approach are equally likely to explain the measured spectrum.

A retrieval code like \texttt{Helios-r2} can be used to study various details of brown dwarf atmospheres, e.g. with already available SpeX or \textit{HST} data - or with future observations provided by the \textit{JWST}.
In that respect, we aim to add more sophisticated cloud parameterizations to the retrieval model in the future. Given the wide wavelength coverage of the \textit{JWST}, it might  be possible to check the validity of the various competing cloud models currently employed by the different atmospheric modeling groups (see, e.g., \citet{Helling2008MNRAS.391.1854H} for an overview). Other possible applications related to brown dwarfs are to test the applicability of the commonly applied modeling approximation that the atmospheres are in chemical equilibrium or to verify the idea of \cite{Tremblin2016ApJ...817L..19T} that the L-T transition is mostly caused by a chemical instability rather than by clouds.

\acknowledgments

D.K., K.H., and M.O. acknowledge partial financial support from the Center for Space and Habitability (CSH), the PlanetS National Center of Competence in Research (NCCR), the Swiss National Science Foundation and the Swiss-based MERAC Foundation.
D.K. would like to thank Nicole Allard for providing her sodium and potassium resonance line-wing profiles and Paul Molli\`{e}re for discussing the details of implementation.

\bibliographystyle{aasjournal}
\bibliography{references}

\appendix
\section{Error inflation posterior}

\begin{figure*}[hb]
	\begin{center}
		\includegraphics[width=0.8\textwidth]{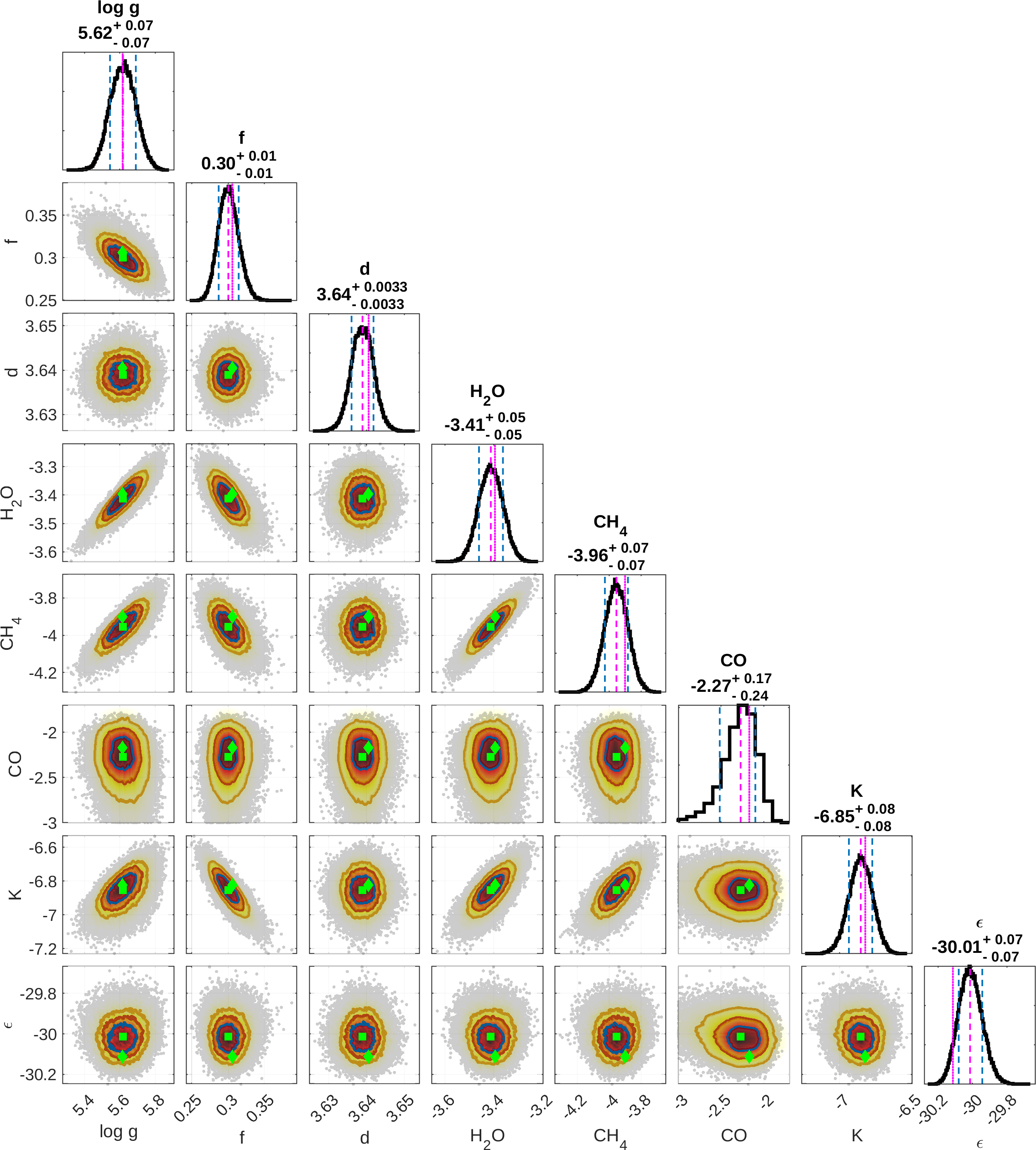}
	\end{center}
	\caption{Posteriors for the free chemistry retrieval of $\epsilon$ Indi Ba including the error inflation parameter. The figure shows the same posteriors as those presented in Figure \ref{fig:epsind_retrieval_fc} but with the error inflation parameter $\epsilon$ added and omitting all unconstrained parameters.}
	\label{fig:epsind_retrieval_fc_with_eps}
\end{figure*}

The retrievals in Sect. \ref{sec:brown_dwarf_retrieval} all include an additional error inflation parameter $\epsilon$, introduced in Sect. \ref{sect:likelhood_error_inflation} that has been omitted from the posterior plots for presentational reasons. To investigate the potential impact of $\epsilon$ on the other retrieval parameters, we show the posteriors of the free chemistry retrieval for Eps Ind Ba including $\epsilon$ in Figure \ref{fig:epsind_retrieval_fc_with_eps}. Note that we omit the unconstrained parameters in this figure. The $\epsilon$ parameter is well constrained and is not correlated with any of the other parameters. The same behavior is also found for all the other retrievals.


\section{Impact of a gray cloud layer}
\label{sect:app_cloud}

\begin{figure*}[ht]
	\begin{center}
		\includegraphics[width=0.9\textwidth]{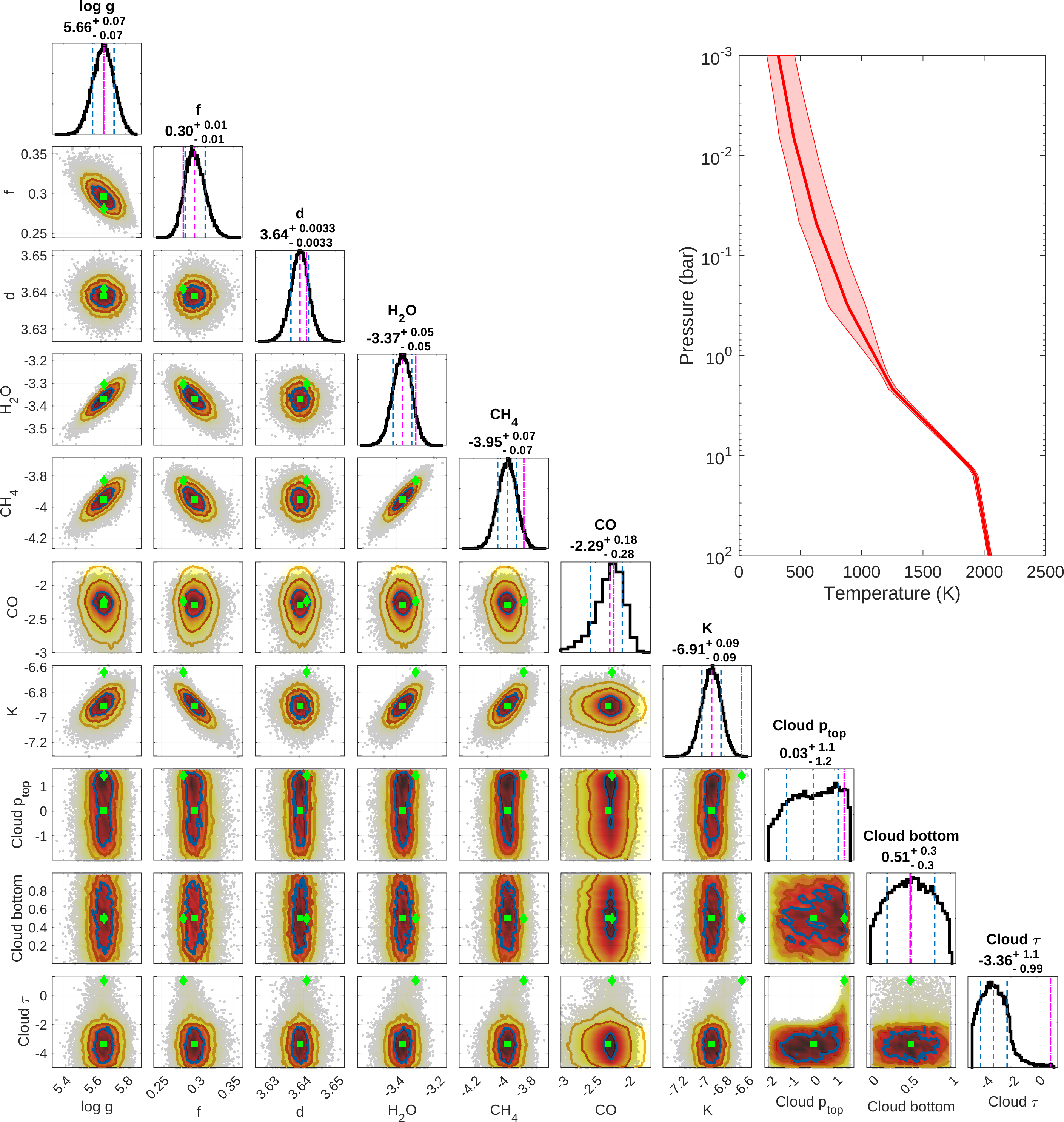}
	\end{center}
	\caption{Posteriors for the free chemistry retrieval of $\epsilon$ Indi Ba including a gray cloud layer. For this figure, we omitted all unconstrained parameters (\ce{NH3}, \ce{CO2}, and \ce{H2S} abundances).}
	\label{fig:epsind_retrieval_fc_cloud}
\end{figure*}

As mentioned in Sect. \ref{sect:discussion}, the shallow temperature profile in the lower atmosphere of Eps Ind Ba we obtain for the equilibrium and free chemistry retrievals might be caused by the neglect of clouds in the forward model. To check this issue, we repeat the free chemistry retrieval but additionally included a gray cloud layer in the forward model. The cloud has three free parameters: the cloud optical depth, the pressure of the cloud layer top, and the bottom of the cloud layer, defined as the fraction of the top pressure. The prior distributions for all three parameters are log-uniform distributions. The corresponding results are shown in Figure \ref{fig:epsind_retrieval_fc_cloud}, where we omit the posteriors for the unconstrained parameters (\ce{NH3}, \ce{CO2}, and \ce{H2S} abundances) for presentational reasons. 

The posterior distributions for the cloud parameters suggest that none of them are constrained. We only obtain an upper limit for the optical depth, while the top and bottom of the cloud layer are prior-dominated. The addition of the cloud layer also has no impact on the other retrieval parameters. We essentially obtain the same posterior distributions as in the cloud-free case (see Figure \ref{fig:epsind_retrieval_fc}), including the very shallow temperature profile in the lower atmosphere. The absence of a gray cloud layer is, thus, unable to explain this peculiar result.


\end{document}